\documentclass{ws-rv961x669}
\usepackage[square]{ws-rv-van}
\usepackage{tikz}
\usetikzlibrary{calc}
\usepackage{graphicx}
\usepackage{amsmath}
\usepackage{slashed}
\usepackage{upgreek}
\usepackage{url}
\usepackage{hyperref}

\usepackage{xcolor}

\def\cropnote{\relax}

\begin{document}

\chapter*{ARS Leptogenesis}
\author[]{M.~Drewes${}^{\ast\dagger}$, B.~Garbrecht${}^{\ast}$, P.~Hern\'andez${}^{\ddagger\S}$\footnote{Corresponding Author.}, M.~Kekic${}^{\ddagger}$, J.~Lopez--Pavon${}^{\S}$, J.~Racker${}^{\P}$, N.~Rius${}^{\ddagger}$, J.~Salvado${}^{\ddagger}$, D.~Teresi${}^{\parallel}$}
\address{${}^{\ast}$Technische Universit\"at M\"unchen, Physik-Department, James-Franck-Stra{\ss}e, 85748 Garching, Germany}
\address{${}^{\dagger}$Centre for Cosmology, Particle Physics and Phenomenology, Universit\'e catholique de Louvain, Louvain-la-Neuve B-1348, Belgium}
\address{${}^\ddagger$Instituto de F\'isica Corpuscular, Universidad de Valencia and CSIC, \\ Edificio Institutos Investigaci\'on, Catedr\'atico Jos\'e Beltr\'an 2, 46980 Spain
%m.pilar.hernandez@uv.es
}
\address{${}^\S$Theory Division, CERN, 1211 Geneve 23}
\address{${}^\P$Instituto  de  Astronom\'{i}a  Te\'{o}rica y Experimental (IATE), Universidad Nacional de C\'{o}rdoba (UNC) - Consejo Nacional de Investigaciones Cient\'{i}ficas y T\'{e}cnicas (CONICET), C\'{o}rdoba, Argentina}
\address{${}^\parallel$ 
Service de Physique Th\'eorique - Universit\'e Libre de Bruxelles, Boulevard du Triomphe, CP225, 1050 Brussels, Belgium}

%\markboth{P. Hern\'andez, M. Kekic, J. Lopez-Pavon, J.~Racker, N.~Rius, J.~Salvado}{ARS Leptogenesis}

%%%%%%%%%%%%%%%%%%%%% Publisher's Area please ignore %%%%%%%%%%%%%%%
%
%\catchline{}{}{}{}{}
%
%%%%%%%%%%%%%%%%%%%%%%%%%%%%%%%%%%%%%%%%%%%%%%%%%%%%%%%%%%%%%%%%%%%%

% %\pdfoutput=1
% \synctex=1
% \NeedsTeXFormat{LaTeX2e}
% \documentclass[12pt]{article}
% 
% \usepackage[reqno]{amsmath}
% 
% \usepackage{mathrsfs}
% \usepackage{amssymb}
% \DeclareMathOperator\arctanh{arctanh}
% 
% \usepackage{geometry}
% \geometry{
%   includefoot,
%   margin=2.3cm
% }
% \usepackage{afterpage}
% \usepackage[sort&compress,numbers,colon,merge]{natbib}
% \bibliographystyle{apsrev}
% \usepackage{subcaption}
% \usepackage{bbm}
% \usepackage{epsfig}
% \usepackage{array}
% \usepackage{float}
% \usepackage{color}
% \usepackage{booktabs}
% 
% \usepackage{graphicx}% Include figure files
% 
% %\parindent 0pt
% \usepackage{cancel}
% 
% \usepackage{wasysym}
% \usepackage{url}
% \usepackage{color}
% 
% \usepackage[colorlinks=true,linkcolor=red,citecolor=green,filecolor=magenta,urlcolor=cyan,bookmarks=true,bookmarksopen=true]{hyperref}
% Some other macros used in the sample text
\def\ra{\rightarrow}
\def\gs{\mathrel{
   \rlap{\raise 0.511ex \hbox{$>$}}{\lower 0.511ex \hbox{$\sim$}}}}
\def\ls{\mathrel{
   \rlap{\raise 0.511ex \hbox{$<$}}{\lower 0.511ex \hbox{$\sim$}}}}

\newcommand{\dms}{\mbox{$\Delta m^2_{\odot}$}}
\newcommand{\dma}{\mbox{$\Delta m^2_{\rm A}$}}
\newcommand{\meff}{\mbox{$\left| m_{ee} \right|$}}
\newcommand{\be}{\begin{eqnarray}}
\newcommand{\ee}{\end{eqnarray}}
\newcommand{\etal}{{\it et al.}}
\def\nue{{\nu_e}}
\def\anue{{\bar\nu_e}}
\def\numu{{\nu_{\mu}}}
\def\anumu{{\bar\nu_{\mu}}}
\def\nutau{{\nu_{\tau}}}
\def\anutau{{\bar\nu_{\tau}}}
\def\lsim{\:\raisebox{-0.5ex}{$\stackrel{\textstyle<}{\sim}$}\:}
\def\gsim{\:\raisebox{-0.5ex}{$\stackrel{\textstyle>}{\sim}$}\:}
\def\ltap{\ \raisebox{-.4ex}{\rlap{$\sim$}} \raisebox{.4ex}{$<$}\ }
\def\gtap{\ \raisebox{-.4ex}{\rlap{$\sim$}} \raisebox{.4ex}{$>$}\ }

\newcommand{\hc}{\ensuremath{\mathrm{h.c.}}}
\newcommand{\Slash}[1]{#1\!\!\!/}
\newcommand{\eps}{\mbox{$\epsilon$}}
\newcommand{\tr}{\ensuremath{\mathrm{Tr}}}
% Some other macros used in the sample text
\def\be{\begin{equation}}
\def\ee{\end{equation}}
\newcommand{\ba}{\begin{array}{c}}
\newcommand{\baz}{\begin{array}{cc}}
\newcommand{\bad}{\begin{array}{ccc}}
\newcommand{\bav}{\begin{array}{cccc}}
\newcommand{\baf}{\begin{array}{ccccc}}
\newcommand{\bena}{\begin{eqnarray}}
\newcommand{\eena}{\end{eqnarray}}
\newcommand{\bea}{\begin{equation} \begin{array}{c}}
\newcommand{\eea}{ \end{array} \end{equation}}
\newcommand{\ea}{\end{array}}
\newcommand{\braket}[1]{\ensuremath{\left\langle#1\right\rangle}}

\def\eps{\epsilon}

\newcommand{\michael}[1]{{\color{blue}#1}}

\newcommand{\Eqref}[1]{Eq.~(\ref{#1})}
\newcommand{\Eqsref}[2]{Eqs.~(\ref{#1},\ref{#2})}
\newcommand{\Figref}[1]{Fig.~\ref{#1}}
\newcommand{\Tabref}[1]{Tab.~\ref{#1}}

\begin{abstract}
We review the current status of the leptogenesis scenario originally
proposed by Akhmedov, Rubakov and Smirnov (ARS). It takes place in the parametric
regime where the right-handed neutrinos are at the electroweak scale or below
and the CP-violating effects are induced by the coherent superposition of
different right-handed mass eigenstates.
Two main theoretical approaches to derive quantum kinetic equations,
the Hamiltonian time evolution as well as the
Closed-Time-Path technique are presented, and we discuss their relations. For scenarios
with two right-handed neutrinos, we chart the viable parameter space. Both, a Bayesian
analysis, that determines the most likely configurations for viable leptogenesis
given different variants of flat priors, and a determination of the maximally allowed
mixing between the light, mostly left-handed, and heavy, mostly right-handed, neutrino
states are discussed.
Rephasing invariants are shown to be a useful tool to classify
and to understand various distinct contributions
to ARS leptogenesis that can dominate in different parametric regimes.
While these analyses are carried out for the parametric regime where initial asymmetries
are generated predominantly from lepton-number conserving, but flavor violating effects,
we also review the contributions from lepton-number violating operators and identify the regions of parameter space
where these are relevant.
\end{abstract}

\newpage

\body

\tableofcontents

\section{Introduction}

One of the most interesting implications of the extensions of the Standard Model (SM) with massive neutrinos is the possibility to explain the baryon asymmetry in the Universe via leptogenesis
\cite{Fukugita:1986hr}. This mechanism has been shown to be robust and generic in seesaw models that involve a very high scale of new physics, much higher that the electroweak scale,  $M_N \gg v$. In the standard scenarios, leptogenesis  takes place during the freeze-out of some heavy states that can decay violating charge-parity CP and  lepton number $L$. These high-scale scenarios have been studied extensively. For comprehensive reviews  see~\cite{Davidson:2008bu} .

 Akhmedov, Rubakov and Smirnov~\cite{Akhmedov:1998qx}  (ARS) studied the possibility to generate a baryon asymmetry  in type I seesaw models at a much lower scale, $M_N \lesssim v$. The key observation is that the small Yukawa couplings required to explain neutrino masses in this low-scale scenario could be 
small enough to ensure
that some of the sterile states [also referred to as right-handed (RH) neutrinos] might not reach thermal equilibrium before the electroweak phase 
transition, when sphaleron processes are switched off. ARS leptogenesis therefore is a freeze-in scenario.
Pending on the parametric regime, lepton-number violating (LNV) processes may be negligible
both in the generation of the asymmetries as well as in the washout because the Majorana mass
of the RH neutrinos is small compared to the temperature.
In that situation, the initial asymmetries in active leptons are purely flavored and lepton-number conserving (LNC), and eventually total asymmetries in the active sector arise and are approximatly counterbalanced by those in the sterile sector. If this situation survives until 
the electroweak phase transition, a net baryon asymmetry results, since the eventual equilibration later on can no longer be transmitted to the baryons in the absence of efficient sphaleron transitions.
In other regions of parameter space LNV contributions may be relevant or even dominating in the
source as well as washout terms and then must be accounted for.
The aim of this chapter is to review the ARS mechanism of leptogenesis.

The model involves the simplest extension of the SM with $n_R$ heavy Majorana singlets
\begin{eqnarray}
{\cal L} = {\cal L}_{SM}+
 \bar{N}_k \, i \slashed\partial \, N_k - \left( \frac 12  \left (M_N\right)_{j k} \bar{N}_j^c N_k + \lambda_{\alpha k} \, \bar{\ell}_\alpha \phi^c N_k + \text{h.c.} \right)
\label{eq:lag}
\end{eqnarray}
where $N_k$ are RH spinors (such that $P_R N_k=N_k$), 
$\lambda$ is a $3\times n_R$ complex matrix, $M_N$ is a $n_R$-dimensional  complex symmetric matrix, and $\phi^c=\epsilon\phi^*$. The spectrum of this theory contains three lighter states with a mass matrix given by the famous seesaw formula
\begin{eqnarray}
M_\nu = -{v^2 \over 2} \lambda M_N^{-1} \lambda^T,  
\end{eqnarray}
where $v=246\,{\rm GeV}$,
and $n_R$ heavy ones with masses of ${\mathcal O}(M_N)$. The naive seesaw scaling (exact for one family) relating Yukawas with the light and heavy masses is therefore
$M_\nu \sim {\mathcal O}({v^2 y^2\over 2 M_N}$), where $y$ is specified through \eqref{lambda:bidiag}
below.

 In most of this work we will assume the minimal scenarios where $n_R=2,3$.  Different parametrizations of the Yukawa matrices have been used in the literature. 
 For some purposes the parametrization in terms of the two unitary matrices that bi-diagonalize $\lambda$ is useful:  
 \begin{eqnarray}
 \label{lambda:bidiag}
 \lambda = V^\dagger {\rm diag}(
y_1,y_2,y_3) W.
 \end{eqnarray}
For the purpose of parameter scanning however the Casas-Ibarra~\cite{Casas:2001sr} parametrization is most appropriate. For $n_R=2$ it reads:
\begin{eqnarray}
\lambda = - i U^*_\nu \sqrt{M_\nu^{\rm diag}}~ P_{NO} ~R^T(z) \sqrt{M_N}{\sqrt{2}\over v},
\label{eq:yci}
\end{eqnarray}
where $U_\nu$ is the PMNS matrix,  $M_\nu^{\rm diag}$ is the diagonal matrix of the light neutrino masses  (note that the lightest neutrino is massless because only two Majorana singlets are included), $M_N\equiv M_N^{\rm diag}={\rm diag}(M_{1},M_{2})$, where $M_{1},M_{2}$ are the heavy neutrino masses and without loss
of generality, we choose the mass basis for the RH neutrinos, $P_{\rm NO}$ is a  $3\times 2$ matrix that depends on the neutrino ordering (NH, IH) 
\begin{eqnarray}
 P_{NH} =\left(\begin{array}{cc} 0& 0 \\
 1 & 0 \\
 0 & 1 \end{array}\right),~ P_{IH} =\left(\begin{array}{cc} 1 & 0\\ 0& 1\\ 0 & 0\end{array}\right),
 \end{eqnarray}
and finally $R(z)$ is a generic two dimensional orthogonal complex matrix that depends on one complex angle $z \equiv \theta + i \gamma$.  For $n_R=3$, the parametrization is 
\begin{eqnarray}
\lambda = - i U^*_\nu \sqrt{M_\nu^{\rm diag}}~ R^T(z_1,z_2,z_3) \sqrt{M_N}{\sqrt{2}\over v},
\end{eqnarray}
where $R(z_1,z_2,z_3)$ is a 3 dimensional complex orthogonal matrix that depends on three complex angles, $z_i, i=1,2,3
$. 

\section{Sakharov Conditions }

As it is well known, three necessary Sakharov conditions need to be met in order to generate a baryon asymmetry from a symmetric initial condition. The model of Eq.~\eqref{eq:lag} satisfies all of them:
\begin{itemize}
\item Baryon number $B$ is violated by sphaleron processes 

\item C and CP violation
\end{itemize}
 
Charge C is violated maximally, while CP violation arises from the complex nature of the flavor parameters $Y$ and $M_N$. After performing all the allowed field rephasings, it is easy to see that 3(6) physical CP-violating phases exist for $n_R = 2(3)$. 
\begin{itemize}

\item Out-of-equilibrium condition
\end{itemize}

 Assuming that the only interactions of the singlets $n_R$ are the Yukawa couplings in Eq.~\eqref{eq:lag}, it is easy to estimate when the states will reach thermal equilibrium.
 The scattering rate of the sterile states at temperature $T$ is roughly  $\Gamma_s(T) \sim y^2 T$, with
 $y$ defined in Eq.~\eqref{lambda:bidiag}. 
 Comparing this to the Hubble expansion rate in a radiation dominated universe, and using the naive seesaw scaling:
 \begin{eqnarray}
{ \Gamma_s(T) \over H(T)} \sim {y^2 M_P^*\over T} \sim {M_\nu M_N M_P^*\over v^2 T} \sim   \left({M_\nu \over 0.05 {\rm eV}}\right)\left({M_N \over 10 \rm GeV }\right) \left({T_{EW} \over T}\right),
 \end{eqnarray}
 where $T_{EW} \sim 140$ GeV is the temperature of the electroweak phase transition,
 $M_P^*=M_{\rm Pl}\sqrt{45/(4\pi^3 g_*)}$, $M_{\rm Pl}$ is the Planck mass, and
 $g_*$ the effective number of relativistic degrees of freedom at the temperature $T$.
This naive estimate shows that the singlet states thermalize quite close to the electroweak phase transition so it is not unnatural to have at least one state that has not reached thermal equilibrium by the time the sphaleron processes switch off.

As it is well known, CP-violating observables such as a  lepton asymmetry, require the quantum 
interference of different amplitudes that involve different  CP-violating phases as well as 
CP conserving ones. While in the standard scenario of leptogenesis from out-of-equilibrium decay 
the CP-conserving phases can be computed from the absorptive parts of the one-loop amplitudes, in the ARS case
they are most suitably described by oscillations among the RH neutrinos.\footnote{Note that in the resonant regime of standard leptogenesis, both descriptions, in terms of oscillations
or absorptive loop contributions can be valid and in the extremely resonant limit,
the oscillation picture can be more suitable. We refer to Refs.~\cite{Garbrecht:2011aw,Garbrecht:2014aga,Dev:2014wsa,Kartavtsev:2015vto} as well as the accompanying review
article on resonant leptogenesis~\cite{leptogenesis:A03} for further details.} 
Indeed the sterile states are produced in flavor combinations that get modified in propagation, because they are a superposition of the mass eigenstates. Two scales are therefore relevant
in the generation of the asymmetry the time when the oscillation rate is similar to the Hubble expansion or oscillation time  
\begin{eqnarray}
t_{\rm osc}(ij) \propto \left( \Delta M_{Nij}^2 M_P^* \right)^{-1/3},
\end{eqnarray}
where $\Delta M_{N\,ij}^2=M_{j}^2-M_{i}^2$,
and the equilibration time when the scattering rate is of the order of the Hubble expansion:
\begin{eqnarray}
 t_{\rm eq}(\alpha) \propto (y^2 M_P^*)^{-1}\sim \left({M_\nu M_N M_P ^*\over v^2}\right)^{-1}.
 \end{eqnarray}
  Unless  the mass eigenstates are extremely degenerate, there is a hierarchy of scales:
\begin{eqnarray}
t_{\rm osc} \ll t_{\rm eq} \sim t_{EW}.
\end{eqnarray}
Note however that there are several flavors and therefore several oscillation and equilibration rates. 

The generation of the asymmetry in the different flavors is effective at $t\sim t_{\rm osc}$, so it takes place at temperatures much higher than the electroweak phase transition. At later times, when oscillations are very fast, quantum effects are no longer possible and the lepton asymmetries no longer grow, but the total lepton asymmetry keeps evolving because the equilibration rate of different flavors is different. After all the states reach the equilibration time $t > t_{\rm eq}$, the asymmetry drops exponentially. If the EW phase transition happens before that, the subsequent evolution of the lepton
asymmetries no longer affects the baryons and therefore whatever baryon asymmetry survives 
until $t_{EW}$ remains. 

\section{CP invariants and naive estimates}
\label{sec:cpinvariants}

CP-violating observables should be flavor basis invariant and be sensitive to the physical phases. In the case where we can consider all the flavor parameters as small so that we can Taylor expand in them, it is easy to find the basis independent invariants that usually require a minimum power of the flavor parameters. 
These invariants are useful because they allow to estimate the size of the asymmetry rather simply.

A well-known example is that of the SM with massless neutrinos. There is only one physical CP-violating phase in the quark sector and the lowest-order 
basis invariant combination of the Yukawas is the famous Jarlskog invariant:
\begin{eqnarray}
I_{\rm CP} &\propto&
 \det\left(-i [\lambda_d \lambda_d^\dagger, \lambda_u \lambda_u^\dagger]\right) \nonumber\\ &\propto 
 & - 2~J
 (m_t^2 \!-\!m_u^2)  (m_t^2 \!-\!m_c^2)  (m_c^2\! -\!m_u^2)  (m_b^2 \!-\!m_s^2)  (m_b^2\! -\!m_d^2)  (m_s^2 \!-\!m_d
^2) , 
\end{eqnarray} 
where $J\equiv {\rm Im}[V_{ij}^* V_{ii}
 V_{ji}^* V_{jj}]$ and $V$ is the CKM matrix. The invariant $I_{CP}$ contains two factors: a rephasing invariant, the $J$ factor, that depends on mixing angles, which 
ensures that the phase becomes unphysical if any of the mixing angle vanishes, and a GIM factor that depends on the quark mass differences, which 
ensures that the same thing happens if any two quarks are degenerate. 
Any flavor-blind CP-violating observable such as the baryon asymmetry, if generated at $T_{EW}$ in electroweak baryogenesis scenarios for example, is expected to be proportional to $I_{CP}$, which fails by many orders of magnitude as is found by direct computation \cite{Gavela:1993ts,Huet:1994jb,Gavela:1994dt}.
 
In the seesaw model the situation is much more complicated because there are three/six independent CP-violating phases for $n_R=2/3$. The flavor or weak basis (WB) invariants are constructed from products of the lepton Yukawa and mass matrices. 
Those  sensitive to the CP-violating phases have been derived in  \cite{Branco:2001pq} within the 
 minimal type I seesaw model:
%In weak basis (WB)  invariants sensitive to the CP violating phases which appear in leptogenesis are derived, within the minimal type I seesaw model.
 all of them should vanish if CP 
is conserved, and conversely the non-vanishing of any of these invariants signals CP violation.
They are invariant under the basis transformations:
\begin{eqnarray}
\label{eq:basis}
\ell & \rightarrow &V_L \ell\,,
\nonumber \\
e_R & \rightarrow &V_R e_R\,, 
\nonumber \\
N & \rightarrow &W_R N \,,
\end{eqnarray}
where $e_R$ are the RH charged leptons of the SM.
In general, there are  3 $\times (n_R-1)$ independent phases, and thus the same number of independent WB invariants.
For $n_R = 3$, three  CP-violating 
WB invariants can be constructed using just  the neutrino Yukawa and mass matrices.
Defining $h=\lambda^\dagger \lambda$, and $H_M=M_N^\dagger M_N$, they can be written as:
\begin{eqnarray}
\label{eq:inv1}
I_1 &\equiv& {\rm Im Tr} [h H_M M_N^* h^* M_N],  \\
\label{eq:inv2}
I_2 &\equiv& {\rm Im Tr} [h H_M^2 M_N^* h^* M_N], \\
\label{eq:inv3}
I_3 &\equiv& {\rm Im Tr} [h H_M^2 M_N^* h^* M_N H_M]  \ .
\end{eqnarray}
Since the $I_a$ are WB invariants, we can evaluate them in any basis. In  the 
WB where the sterile neutrino mass matrix $M_N$ is real and diagonal ($M_N={\rm Diag}(M_{1},\cdots,M_{n_R})$), one obtains:
\begin{equation}
I_a = \sum_j  {\rm Im}(h_{ij}^2) g_a(M_{i},M_{j}) \ ,
\end{equation}
with $g_1(M_{i},M_{j}) = M_{i} M_{j}^3$, $g_2(M_{i},M_{j}) = M_{i} M_{j}^5$  and $g_3(M_{i},M_{j}) = M_{i}^3 M_{j}^5$.
From the above equations one can see that 
in the case $n_R = 2$ there is only one independent CP-violating invariant, which at lowest order in the sterile neutrino masses is $I_1$.
Note that the quantities 
${\rm Im} (h_{ij}^2) =  {\rm Im} [(\lambda^\dagger \lambda)_{ij}^2]$ are related to the CP asymmetries which appear in 
unflavored leptogenesis via heavy Majorana neutrino decay, 
\begin{equation}
\epsilon_i \propto 
%\sum_j  {\rm Im} (h_{ij}^2)  \, f(M_{Ni},M_{Nj}) = 
\sum_j {\rm Im} [(\lambda^\dagger \lambda)_{ij}^2] \, f(M_{i},M_{j}) \ , 
\end{equation}
where $f$ is a loop a function.

 The remaining independent WB invariants that can be constructed  involve  also the charged lepton Yukawa
couplings, $\lambda_\ell$.
Since ARS leptogenesis occurs at $T \gg M$, CP-violating phases associated with the Majorana character of the sterile neutrinos are suppressed by $M/T \ll 1$ and do not play any role, if these effects are neglected.
In this case, the appropriate CP-violating invariants at lowest order for $n_R = 3$ are (only the first one for $n_R=2$): 
\begin{eqnarray}
\label{eq:inv4}
\bar{I}_1' &\equiv& {\rm Im Tr} [h H_M^2 \bar h H_M],  \\
\label{eq:inv5}
\bar{I}_2' &\equiv& {\rm Im Tr} [h H_M^3 \bar h H_M ],  \\
 \label{eq:inv6}
\bar{I}_3' &\equiv& {\rm Im Tr} [h H_M^3 \bar h  H_M^2],  
\end{eqnarray}
where $\bar h \equiv \lambda^\dagger \lambda_\ell \lambda_\ell^\dagger \lambda$, and in the same WB as before, 
 \begin{equation}
\bar{I}_a' = \sum_j  {\rm Im}(h_{ij} \bar{h}_{ji} ) g_a'(M_{i},M_{j}) \ ,
\end{equation}
with $g_1'(M_{i},M_{j}) = M_{i}^2 M_{j}^4$, $g_2'(M_{i},M_{j}) = M_{i}^2 M_{j}^6$, $g_3'(M_{i},M_{j}) = M_{i}^4 M_{j}^6$ and
\begin{equation}
\label{eq:CPLC}
 {\rm Im}(h_{ij} \bar{h}_{ji} ) = 
 \sum_{\alpha} \lambda_\alpha^2 \, {\rm Im} [\lambda_{\alpha i} \lambda^*_{\alpha j}  (\lambda^\dagger \lambda)_{ij}] \ .
 %\sum_{\alpha, \beta} \lambda_\alpha^2 \, {\rm Im} [Y_{\alpha i} Y^*_{\alpha j}  Y_{\beta j} Y^*_{\beta i}] \ .
\end{equation}
In the case $n_R=2$, we find the third independent WB invariant  at higher order in $\lambda$, 
$ {\rm Im Tr} [h^2 H_M^2  \bar h  H_M] $.

One can also construct WB invariants involving $\lambda_\ell$ and sensitive to Majorana phases, related to  
the flavored CP asymmetries in heavy neutrino decays, by  just changing the matrix $h$ by $\bar{h}$ in 
Eqs.~\eqref{eq:inv1}--\eqref{eq:inv3}~\cite{Branco:2001pq}, but they contain the same phases as the ones described here.

Note however that in ARS leptogenesis (and more generally in  flavored leptogenesis), charged lepton
Yukawa couplings are not small parameters: in fact, they mediate fast interactions that are
resummed in the Boltzmann equations, leading to density matrices diagonal in
the charged lepton mass basis. Therefore for the relevant CP-violating observables a
perturbative expansion in the charged lepton Yukawas does not work, but rather a 
projection on the different flavors: the total lepton asymmetry
is an incoherent sum of the flavor lepton asymmetries, which evolve independently.
As a consequence the WB invariants $ \bar{I}_i$ do not appear in the final result, 
but the structure of the flavored CP-asymmetries is dictated by Eq.~(\ref{eq:CPLC}):
for instance the ($L$-conserving) CP asymmetry in flavor $\alpha$ produced in the decay of the sterile neutrino $i$
is (see the accompanying article~\cite{leptogenesis:A01}) 
\begin{equation}
\label{eq:ELC}
\epsilon_{i \alpha} \propto 
 \sum_j {\rm Im} [  (\lambda^\dagger P_\alpha \lambda)_{ji} (\lambda^\dagger \lambda)_{ij} ] \,  f(M_{i},M_{j}) = 
%[ Y^\dagger Y Y^\dagger I_{\alpha} Y ]   =
 \sum_j {\rm Im} [ \lambda_{\alpha i}  \lambda^*_{\alpha j}  (\lambda^\dagger \lambda)_{ij}] \, f(M_{i},M_{j})  \ ,
\end{equation}
where $P_\alpha$ is the projector on flavor $\alpha$;
and in ARS leptogenesis, where all  sterile neutrino species contribute, one finds 
\begin{equation}
\label{eq:CP_ARS}
\Delta_\alpha = 
 \sum_{i,j} {\rm Im} [ \lambda_{\alpha i}  \lambda^*_{\alpha j}  (\lambda^\dagger \lambda)_{ij}] \,  \tilde{f}(M_{i},M_{j})  \ .
\end{equation}
 We can still use the above WB invariants to estimate the size of the baryon 
asymmetry, in particular its dependence on the neutrino Yukawa couplings, $\lambda$,
in processes like ARS leptogenesis which are flavor blind.
% when they can be considered as small parameters and a perturbative expansion is expected to give the correct result, i.e., in the weak washout limit.}
From the above equations, we see  that it must be at least fourth order in $\lambda$, 
however it turns out that $\sum_\alpha \Delta_{\alpha} = 0$,
 therefore 
 %the flavor blind asymmetry, i.e., 
  the baryon number asymmetry, which is proportional to 
 the sum of all the flavored asymmetries, should be ${\cal O}(\lambda^6)$. 
 Thus the final  asymmetry in ARS is expected to have the form
 \begin{equation}
 \label{eq:YB}
 Y_{\Delta B} \propto \sum_\alpha (\lambda \lambda^\dagger)_{\alpha \alpha} \Delta_\alpha \ .
 \end{equation}
%and as we will see, this is indeed the case.
This reflects the fact that the CP asymmetries $\Delta_{\alpha}$ (and $\epsilon_{i \alpha}$)  are lepton number conserving:
 a net lepton number asymmetry is generated only because each flavor lepton asymmetry evolves 
 at a different rate. Notice that in the case of heavy Majorana neutrino decays, in general there is an additional piece of the CP-asymmetry, 
 \begin{equation}
 \label{eq:ELV}
 \bar{\epsilon}_{i \alpha} \propto   \sum_j {\rm Im} [ \lambda^*_{\alpha i}  \lambda_{\alpha j}  (\lambda^\dagger \lambda)_{ij}] 
 \, g(M_{i},M_{j})
 \end{equation}
 which violates lepton number and does not vanish when summed over all flavors.
 Only in purely flavored leptogenesis, or in inverse seesaw scenarios where 
 $\bar{\epsilon}_{i \alpha} \ll \epsilon_{i \alpha}$, a similar cancellation occurs.

 However the dependence of the baryon asymmetry on $\lambda$ may change if one of the lepton flavors $\alpha$ is almost decoupled 
when the sphalerons freeze out, while the others have equilibrated; in this case the Yukawa couplings of the 
fast interacting species are not small parameters. 
Then,  by lepton number conservation
 the asymmetry sequestered in the flavor $\alpha$ is equal and opposite to the asymmetry in the rest of 
 the leptonic sector, which is partially transformed into the baryon asymmetry. In this situation, the 
 final result is ${\cal O}(\lambda^4)$.

Even if the final asymmetry is not WB invariant, 
in any basis there is still the freedom to make phase rotations of the lepton 
fields, and the physical quantities must be invariant under such rephasings. 
As a consequence, the CP-violating observables can be written in terms of 
 independent rephasing invariants, as described in  \cite{Jenkins:2007ip}.
In order to find these, it is convenient to write explicitly the unitary matrices $V,W$ 
with the 3 $\times (n_R-1)$ phases that remain physical after all allowed 
field rephasing transformations. For $n_R =3$ we have 
\begin{eqnarray}
W&=&U(\theta_{12}, \theta_{13}, \theta_{23},\delta)^\dagger {\rm Diag}(1, e^{i\alpha_1},  e^{i\alpha_2}) 
\\
V&=&  {\rm Diag}(1, e^{i\phi_1},  e^{i\phi_2}) U(\theta_{12}', \theta_{13}', \theta_{23}',\delta')  \ ,
\end{eqnarray}
where $U(\theta_i, \delta_i)$ represents the standard parameterization of a 3$\times$3 mixing matrix, with 
3 angles and 1 phase.
Notice that the phases $\phi_i$ only enter in processes which depend on both, $V$ and $W$, involving 
$\nu$, $N$ and charged leptons; flavored and ARS leptogenesis are an example of such 
processes, while unflavored leptogenesis depends only on $\lambda^\dagger \lambda$, and therefore on  
$W$. 
%For $n_R=2$, there is only one (Majorana) phase  in $W$ and two in $V$ ($\delta',\phi_1$).

The Majorana phases of $W$, $\alpha_{1,2}$ are determined by two independent invariants of the 
form $ {\rm Im} [(W_{\alpha j} W_{\alpha i}^*)^2]$, while the Jarlskog invariant 
$J_W = - {\rm Im} [W_{23}^* W_{22} W_{32}^* W_{33}]$ determines the phase $\delta$. 
For $n_R=2$, there is a single Majorana phase in $W$, given by the rephasing invariant $ {\rm Im} [(W_{12} W_{11}^*)^2]$.
When considering the parametric regime of ARS leptogenesis where the Majorana nature of the sterile neutrinos does not play a role, only the Dirac phase $\delta$ will be relevant and therefore we expect to find just the Jarlskog invariant of the matrix W, only in the case $n_R=3$.

The remaining invariants involve also the matrix $V$. 
They can be chosen as the Jarlskog invariant $J_V$  and two combinations of the form
${\rm Im} [W^*_{1 i} V_{1 \alpha} V^*_{2 \alpha}  W_{2 i } ]$ for two reference values of $i, \alpha$, 
which fix all the phases in the matrix $V$ ($\delta', \phi_1, \phi_2$).
If $n_R =2$, there are only two more independent phases ($\delta',\phi_1$), and therefore  two 
independent rephasing invariants.

We can now write  the CP asymmetries that appear in ARS leptogenesis in terms of the rephasing invariants as:
 \begin{equation}
  \label{eq:ALC}
 {\rm Im} [(\lambda^\dagger P_\alpha \lambda)_{ji}  (\lambda^\dagger \lambda)_{ij} ]  =
 \sum_{\beta,\delta,\sigma} y_\beta y_\delta y_\sigma^2  \, {\rm Im} 
[W^*_{\beta i} V_{\beta \alpha} V^*_{\delta \alpha}  W_{\delta j} W_{\sigma i}   W^*_{\sigma j}  ]  \ .
\end{equation}
In turn, 
all these invariants  can be written in terms of the 6 (3) independent ones for $n_R = 3(2)$
by using the unitarity of the mixing matrices 
$V, W$.

Since the Majorana phases of $W$ do not contribute in the limit of small sterile neutrino Majorana masses that we are considering, we are left with 4 (2) invariants for $n_R=3(2)$.
The result can be further simplified using that one of the sterile neutrinos is very weakly coupled, so  
in the approximation of neglecting $y_3 \ll y_1,y_2$ 
we obtain:
\begin{eqnarray}
% {\rm Im}(h_{ij} \bar{h}_{ji} )
 {\rm Im}   [ \lambda_{\alpha i}  \lambda^*_{\alpha j} (\lambda^\dagger \lambda)_{ij} ]
&=& y_1^2 y_2^2  (|V_{2\alpha}|^2 - |V_{1\alpha}|^2)
{\rm Im} [ W^*_{1i} W_{1j} W^*_{2j} W_{2i}] 
\nonumber \\
 &+& y_1 y_2 
 \left\{ 
 \left [ y_2^2 |W_{2i}|^2  - y_1^2 |W_{1i}|^2 \right]  
 {\rm Im} [W^*_{1 j} V_{1 \alpha} V^*_{2 \alpha}  W_{2 j} ] \right.
 \nonumber \\
 \label{eq:invLC}
&+& \left. \left[  y_1^2 |W_{1j}|^2 \ - y_2^2 |W_{2j}|^2  \right] 
{\rm Im} [W^*_{1 i} V_{1 \alpha} V^*_{2 \alpha}  W_{2 i } ] \right \} \ .
\end{eqnarray}
Moreover, it can be shown that in this limit the phase 
$\phi_2$ of the matrix $V$ does not appear in Eq.~\eqref{eq:invLC}, thus  only three 
independent invariants contribute for $n_R=3$, which we can choose as:
\begin{eqnarray} 
I_1^{(2)}& =&  - {\rm Im} [W_{1 2}^* V_{1 1} V_{2 1}^*  W_{2 2 } ] ,
\\
I_1^{(3)} &=& {\rm Im} [W_{1 2}^* V_{1 3} V_{2 3}^*  W_{2 2 } ],
%\simeq \theta_{12} \bar{\theta}_{13}  \bar{\theta}_{23} \sin(\delta_L + \phi_1)
\\
I_2^{(3)} &=&  {\rm Im} [W_{13}^* V_{12} V_{2 2}^*  W_{2 3 } ]   \ .
%\simeq  \bar{\theta}_{12}   \theta_{13}  \theta_{23} \sin (\delta -\phi_1)  
 \end{eqnarray}

Although $J_W$ can be  related to the above invariants, 
% by 
%\be
  %{\rm Im} [W^*_{1 3} V_{1 \alpha} V^*_{2 \alpha}  W_{2 3 } ] 
%= \frac{{\rm Im} [(W^*_{1 2} V_{1 \alpha} V^*_{2 \alpha}  W_{2 2 } )
%( W_{23}^* W_{22} W_{32}^* W_{33}) ]  }
%(V^*_{2 \beta} V_{2 \alpha} V^*_{1 \alpha}  V_{1 \beta } )
%{|W_{1 2} W_{2 2} | ^2}   \ , 
%\end{equation}
it is simpler to write the final baryon asymmetry for $n_R=3$ in terms of the four 
rephasing invariants $\{I_1^{(2)},I_1^{(3)},I_2^{(3)},J_W\}$.
In the case $n_R=2$,  only the two independent invariants,  $\{I_1^{(2)},I_1^{(3)}\}$, appear.

\section{ Quantum kinetic equations}
\label{sec:kinetic}

The generation of lepton asymmetries is a purely quantum phenomenon and will therefore be missed
in any classical treatment of the production of the sterile species. A related physical problem is that 
of oscillating neutrinos in the early Universe for which a formalism based on quantum kinetic equations 
was developed a long time ago \cite{Sigl:1992fn}. Other approaches have been proposed \cite{Garbrecht:2011aw,Garny:2011hg}, and while
there is not yet a fully unified formulation, different methods seem to give similar results. 

There are two basic ingredients in any formulation of the problem: the dispersion relation of sterile neutrinos 
in a thermal plasma or refractive index and the scattering rates. 

\subsection{Dispersion relations of sterile neutrinos in a thermal plasma}
\label{sec:dispersive}

The relativistic RH neutrinos acquire dispersive corrections
from the thermal loop made up of Higgs and lepton doublets that
can become of quantitative importance in certain parametric regions. This is in particular the
case for strong mass degeneracy among the RH neutrinos or large active-sterile
mixing, i.e. in the overdamped regime (cf. \sref{sec:overdamped} and \sref{sec:largemix}), where
the damping time scale (or the scale associated with scattering rates)
is of the same order or faster than the time scale of oscillations.
This well-known refractive effect is calculated in the general
context of fermions at finite temperature in Ref.~\cite{Weldon:1982bn}.
Away from the mass-degenerate regime, the thermal mass corrections
appear at next-to-leading order and are therefore subdominant. For leptogenesis
in the strong washout regime, this aspect is discussed in the accompanying
article~\cite{leptogenesis:A04}.

\subsection{Collision rates}
\label{sec:collisionrates}

For ultrarelativistic RH neutrinos, the phase space for reactions
$N\leftrightarrow \ell \phi$ is suppressed at leading order and
only opens up when Standard Model interactions, in particular mediated
by gauge and top-quark Yukawa couplings are included. The leading
logarithmic correction turns out to originate from the $t$-channel
exchange of a doublet lepton $\ell$ in a $2\leftrightarrow2$ process with
gauge radiation. Beyond the leading logarithm, also the leading order
effects from hard scatterings as well as dispersive thermal effects
have been computed.

The production of ultrarelativistic RH neutrinos has first been calculated using the imaginary-time formalism
of thermal field theory in Refs.~\cite{Anisimov:2010gy,Besak:2012qm}. The derivation in the Closed-Time-Path
formalism can be found in Ref.~\cite{Garbrecht:2013urw}. While it appears not to be of immediate consequence for ARS leptogenesis in the mass-range considered here, the interpolation between the production
of ultrarelativistic and non-relativistic RH neutrinos is addressed in Refs.~\cite{Garbrecht:2013gd,Ghisoiu:2014ena}.
More important may be the effects on the rate from the slowly evolving Higgs field
expectation value through the electroweak crossover that is computed in Ref.~\cite{Ghiglieri:2016xye},
but this has not yet been implemented in a phenomenological calculation on ARS leptogenesis.
In the accompanying article~\cite{leptogenesis:A04}, a detailed discussion of these
matters is provided in Sec.~3.2.

\subsection{Raffelt-Sigl formalism}
\label{sec:operator:formalism}

The starting point of this formalism is to consider the time evolution of number density matrices represented
by the expectation values of the number density operators for the particle states (right-helicity states) and antiparticles 
(left-helicity states):
\begin{align}
\langle a_j^\dagger({\mathbf k}) a_i({\mathbf k}')\rangle_T \equiv& (2 \pi)^3 \delta^3({\mathbf k}-{\mathbf k}') \left(\rho(k)\right)_{ij},\nonumber\\
\langle b_i^\dagger({\mathbf k}) b_j({\mathbf k}')\rangle_T \equiv& (2 \pi)^3 \delta^3({\mathbf k}-{\mathbf k}') \left(\bar{\rho}( 
k)\right)_{ij}.
\end{align} 
It is assumed that $M/T$ effects are negligible.
The diagonal elements of $(\rho(k))_{ii}, (\bar{\rho}(k))_{ii}$ represent therefore the number density of the $i$-th sterile particles with positive, negative helicity.  
By working out the time evolution of the number density operators at second order in perturbation theory in the Yukawa interaction, it can be shown\cite{Sigl:1992fn} that the densities satisfy  
\begin{eqnarray}
{d \rho_N \over d t} = -i [H, \rho_N] -{1 \over 2} \left\{ \Gamma^a_N, \rho_N \right\} + {1 \over 2} \left\{ \Gamma^p_N, 1-\rho_N \right\}, 
\end{eqnarray}
where $\Gamma^a_N(k)$ and $\Gamma^p_N(k)$ are the annihilation and production rates of the sterile neutrinos, and 
\begin{eqnarray}
H \equiv {M_N^2 \over 2 k_0} +  {T^2 \over 8 k_0} \lambda^\dagger \lambda,
\end{eqnarray}
includes the refractive effects \cite{Weldon:1982bn} of neutrino propagation in the thermal plasma (we have excluded those effects that are flavor blind, i.e. proportional to the identity
in flavor which drop from the commutator). 

The scattering rates can be written  as
\begin{eqnarray}
\Gamma^{p}_{N ij} &=& \lambda^\dagger_{i\alpha} \rho_F\left({k_0\over T}- \mu_\alpha\right) \gamma_{N }(k,\mu_\alpha) \lambda_{\alpha j},\nonumber\\
\Gamma^{a}_{N ij} &=& \lambda^\dagger_{i\alpha}
\left(1-\rho_F\left({k_0\over T}- \mu_\alpha\right)\right) \gamma_{N
}(k,\mu_\alpha) \lambda_{\alpha j} ,
\end{eqnarray}
where $\rho_F(y) =( \exp y + 1)^{-1}$ is the Fermi-Dirac distribution and $\mu_\alpha$ is the leptonic chemical potential normalized by the temperature.  $\gamma_N$ contain the contributions from all  $2 \rightarrow 2$ processes that produce an $N$:
\begin{eqnarray}
\bar{Q} t \rightarrow \bar{\ell} N;\; t \ell \rightarrow Q N;\; \bar{Q} \ell
\rightarrow \bar{t} N;\; W \ell \rightarrow \bar{\phi} N;\; \ell \phi \rightarrow
W N;\; W \phi \rightarrow \bar{\ell} N ,
\end{eqnarray}  
and $1\leftrightarrow 2$ processes: $\phi \rightarrow \bar{\ell} N$ including resummed soft-gauge interactions. 
All these contributions have been computed for vanishing leptonic chemical potential in \cite{Asaka:2011wq,Besak:2012qm,Ghisoiu:2014ena} and including the effect of a leptonic chemical potential  to linear order in \cite{Hernandez:2016kel} 
.
 
Approximating
\begin{eqnarray}
\gamma_N(k, \mu_\alpha) \simeq \gamma^{(0)}_N(k) + \gamma^{(2)}_N(k) \mu_\alpha, ~~\gamma_N^{(1)} \equiv  \gamma_N^{(2)} -{\rho'_F \over \rho_F} \gamma_N^{(0)},
\end{eqnarray}
with $\rho'_F(y) \equiv {d \rho_F(y)\over dy}$, and inserting these functions in the kinetic equation we get:
\begin{eqnarray}
\label{QB:Valencia}
{d \rho_N \over d t} &=& -i [H, \rho_N] -{ \gamma_N^{(0)}\over 2} \left\{ \lambda^\dagger \lambda  , \rho_N-\rho_F \right\} + \gamma_N^{(1)}  \rho_F  \lambda^\dagger \mu \lambda 
- { \gamma_N^{(2)}\over 2} \left\{ Y^\dagger \mu Y  , \rho_N\right\}, \nonumber\\
{d \bar{\rho}_N \over d t} &=& -i [H^*, \bar{\rho}_N] -{ \gamma_N^{(0)}\over 2} \left\{ \lambda^T \lambda^*  , \bar{\rho}_N-\rho_F \right\} - \gamma_N^{(1)}  \rho_F  \lambda^T \mu \lambda^* +
 { \gamma_N^{(2)}\over 2} \left\{ \lambda^T \mu \lambda^*  , {\bar \rho}_N\right\},\nonumber\\
\label{eq:rhon}
 \end{eqnarray}
where $\mu \equiv {\rm Diag}(\mu_\alpha)$. The coefficients $\gamma^{(i)}_N$, which are functions of $k_0/T$,
 can be found in  \cite{Hernandez:2016kel} .

Finally we need the equations that describe the evolution of the leptonic chemical potentials. These are obtained from the equation 
that describes the evolution of the conserved charges in the absence of neutrino Yukawas, that is the ${B \over 3} -L_\alpha$ numbers, where $L_\alpha$ is the lepton number in the flavor $\alpha$. 
These numbers can only be changed by the same out of equilibrium processes that produce the sterile neutrinos and 
it is possible to relate the integrated rates of these equations to those of the sterile neutrinos and their densities:
\begin{eqnarray}
\dot{n}_{B/3-L_\alpha} & = & 
-2 \int_{k}  \left\{{\gamma_N^{(0)}\over 2} (\lambda \rho_N \lambda^\dagger- \lambda^* \rho_{\bar N} \lambda^T)_{\alpha\alpha} \right.\nonumber\\
&+&\left.\mu_\alpha \left({\gamma_N^{(2)}\over 2} (\lambda \rho_N \lambda^\dagger+\lambda^* \rho_{\bar N} \lambda^T)_{\alpha\alpha} - \gamma_N^{(1)} {\rm Tr}[\lambda\lambda^\dagger P_\alpha] \rho_{F}  \right)\right\},
\label{eq:bml}
\end{eqnarray}
where $P_\alpha$ is the projector on flavor $\alpha$. 

The relation between the leptonic chemical potentials and the approximately conserved charges, $B/3 - L_\alpha$, is given for $T \leq 10^6$ GeV by \cite{Nardi:2006fx}
\begin{eqnarray}
\label{linrel:spectators}
\mu_\alpha = -\sum_\beta C_{\alpha\beta} \mu_{B/3-L_\beta}, \;\; C_{\alpha\beta} = {1\over 711} \left(\begin{array}{ccc} 221 & -16 & -16\\
-16 & 221 & -16 \\
-16 & -16 & 221\end{array}\right),
\label{eq:chem}
\end{eqnarray}
where we have defined $\mu_{B/3-L_\beta}$ by the relation:
\begin{eqnarray}
n_{B/3-L_\alpha} \equiv -2 \mu_{B/3 -L_\alpha} \int_k \rho'_F = {1\over 6} \mu_{B/3 -L_\alpha} T^3.
\end{eqnarray}

Introducing finally the expansion of the Universe and changing variables to the scale factor $x=a$ and $y=k a$, the time derivative of the distribution functions changes to:
\begin{eqnarray}
{d\rho_N(T,k) \over d t} \rightarrow \left. x H_u(x) {\partial \rho_N(x,y)\over\partial x}\right|_{\rm y ~fixed},\nonumber\\
{d n_{B/3 -L_\alpha}\over d t} \rightarrow -2 x H_u(x) {d\mu_{B/3 - L_\alpha}\over d x} \int_k \rho'_F,
\end{eqnarray}
where $H_u(x)%\sqrt{4 \pi^3 g_*(T)\over 45} {T^2\over M_{\rm P}} \equiv
={T^2 \over M_P^*}$ is the Hubble expansion parameter. Assuming a radiation dominated universe with constant number of relativistic degrees of freedom $g_*(T_0)\simeq 106.75$ for $T_0 \geq T_{EW}$, then $x T$= constant that we can fix to one. 

To simplify the numerical solution of these equations,  it is common to consider momentum-averaged equations with the approximation $\rho_N(x,y)= r_N(x) \rho_F(y)$, so that momentum can be averaged:
\begin{eqnarray}
x H_u {d r_N\over d x} &=& -i [\langle H\rangle, r_N]  -{\langle\gamma^{(0)}_N\rangle\over 2} \{\lambda^\dagger \lambda, r_N-1\}
+  \langle \gamma_N^{(1)} \rangle \lambda^\dagger \mu \lambda\nonumber\\   &-&  {\langle \gamma_N^{(2)}\rangle \over 2}  \big\{\lambda^\dagger \mu \lambda,r_N\big\},\nonumber\\ 
x H_u {d r_{\bar N}\over d x} &=& -i [\langle H^*\rangle, r_{\bar N}]  -{\langle\gamma^{(0)}_N\rangle\over 2} \{\lambda^T \lambda^*, r_{\bar N}-1\} 
-  \langle \gamma_N^{(1)} \rangle \lambda^T \mu \lambda^*\nonumber\\   &+&  {\langle \gamma_N^{(2)}\rangle \over 2}  \big\{\lambda^T \mu \lambda^*,r_{\bar N}\big\},\nonumber\\  
x H_u {d{\mu}_{B/3-L_\alpha}\over d x} & = & {\int_k \rho_F\over \int_k \rho'_F}  \left\{{\langle \gamma_N^{(0)}\rangle\over 2} (\lambda r_N \lambda^\dagger- \lambda^* r_{\bar N} \lambda^T)_{\alpha\alpha} \right.\nonumber\\
&+&\left.\mu_\alpha \left({\langle\gamma_N^{(2)}\rangle\over 2} (\lambda r_N \lambda^\dagger+\lambda^* r_{\bar N} \lambda^T)_{\alpha\alpha} - \langle\gamma_N^{(1)}\rangle {\rm Tr}[\lambda\lambda^\dagger P_\alpha]   \right)\right\}.
\label{eq:rhonrhonbarav}
\end{eqnarray}
The averaged rates, $\langle \gamma^{(n)}_N\rangle$ can be found in \cite{Hernandez:2016kel} . 

The baryon to entropy ratio is given by:
\begin{eqnarray}
Y_{\Delta B} = 1.3 \cdot 10^{-3} \cdot \sum_{\alpha} \mu_{B/3-L_\alpha}. 
\end{eqnarray}

It is interesting 
to compare these equations to those used in the literature. In the seminal ARS paper \cite{Akhmedov:1998qx}, only top-quark scatterings were included, and the evolution of the leptonic chemical potentials was neglected. The latter approximation turns out to be too restrictive as an asymmetry builds up in the sterile sector only if at least three sterile species are involved. This approximation missed the important physical effect that the plasma at the relevant temperatures erases any quantum coherence in the charged leptons, which are in kinetic equilibrium also via their Yukawa couplings. The plasma is able therefore to {\it select} a charged-lepton flavor.  
This important point was addressed by Asaka and Shaposhnikov\cite{Asaka:2005pn,Shaposhnikov:2008pf}. They took into account the evolution of the leptonic chemical potentials and  demonstrated that lepton asymmetries could arise also in the minimal case with $n_R=2$.  Compared to Eq.~(\ref{eq:rhon}) and Eq.~(\ref{eq:bml}),
the equations of Refs.~\cite{Asaka:2005pn,Shaposhnikov:2008pf,Canetti:2012zc, Canetti:2012kh}  included only top-quark scatterings,  assumed Boltzmann statistics, 
neglected spectator effects and all non-linear terms in the equations, in particular those of ${\mathcal O}(\mu 
\rho_N)$. The latter approximation was relaxed in Refs.~\cite{Asaka:2011wq,Abada:2015rta,Hernandez:2015wna}.

Finally, in Ref.~\cite{Shuve:2014zua}, the Boltzmann approximation was kept but all relevant scatterings were included in the 
$\mu$ independent terms (and approximately for the $\mu$ dependent ones),  and spectator effects were properly included by considering the evolution of the $B/3-L_\alpha
$ chemical potentials.

\subsection{Closed-Time-Path formalism}
\label{sec:non-equlibrium}

Provided the initial state is known, there are exact equations that in principle describe
the microscopic evolution of an ensemble of non-equilibrium systems.
In a classical theory, these are given by the Liouville
equation. In quantum field theory, one may choose between
an operator formalism leading to von Neumann equations, as reviewed in \sref{sec:operator:formalism},
or a functional approach that we discuss in the present section. The functional
approach leads in a direct manner to
Schwinger-Dyson equations on the Closed-Time-Path (CTP)~\cite{Calzetta:1986cq}.
% As they are formulated in terms of quantum correlation functions, they may
% be viewed as the quantum analogues of the Liouville equations
% for probability density functions.
For practical reasons,
the non-equilibrium reactions that are the essence of particle cosmology
are often described by supplementing cross sections into classical
Boltzmann equations, thus bypassing the first step of the program
shown in \fref{fig:fundamental-to-fluid}. In contrast to this
practice, the ARS scenario is mostly treated in the operator or the CTP approach from the outset
because at its core is the time evolution of \emph{quantum} correlations
among the different RH neutrinos. As indicated in  \fref{fig:fundamental-to-fluid},
further simplifications, ideally by applying controlled approximations, are
necessary in order to make analytical or numerically fast predictions that are suitable for
phenomenological studies.

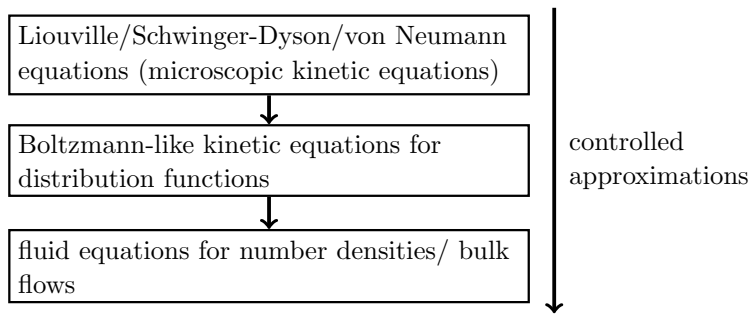
\begin{figure}[ht]
\hskip2.8cm
\parbox{8cm}{
\begin{tikzpicture}
\node [xshift=0cm,yshift=0cm,draw=black,line width=.3mm,fill=white](a){\parbox{6.6cm}{\raggedright\textcolor{black}{ Liouville/Schwinger-Dyson/von Neumann equations (microscopic kinetic equations)}}};

\node [xshift=0cm,yshift=-1.4cm,draw=black,line width=.3mm,fill=white](b){\parbox{6.6cm}{\raggedright\textcolor{black}{ Boltzmann-like kinetic equations for distribution functions}}};

\node [xshift=0cm,yshift=-2.8cm,draw=black,line width=.3mm,fill=white](c){\parbox{6.6cm}{\raggedright\textcolor{black}{ fluid equations for number densities/ bulk flows}}};

\draw [->,draw=black,line width=.5mm] (a)--(b);
\draw [->,draw=black,line width=.5mm] (b)--(c);

\draw [->,draw=black,line width=.5mm] ($(a)!3.8cm!(6,1.0)$)--($(c)!3.8cm!(6,-3.8)$);

\node [xshift=5.2cm,yshift=-1.4cm](c){\parbox{2.5cm}{\raggedright\textcolor{black}{ controlled approximations}}};

\end{tikzpicture}
}

\caption{Path of deriving fluid equations for phenomenological calculations from first principles.\label{fig:fundamental-to-fluid}}
\end{figure}
Both, operator and functional approach, start from first principles, and there is apparent benefit
in verifying results using these complementary methods. Interesting features of the CTP approach
that we mention before going into the technicalities are:
\begin{itemize}
\item
The functional approach makes no reference to the interaction picture states
that are a necessary ingredient to the operator formalism. Rather, the state of the
system can be encoded in two-point functions. This can be an advantage because
in a realistic system, it is non-trivial to relate the interaction picture states
to the physical initial conditions, which are non-Gau{\ss}ian in general~\cite{Garny:2015oza}.
However, for leptogenesis as a weakly coupled model,
this is hardly an issue because the non-Gau{\ss}ian correlations
can be reliably constructed using the straightforward perturbation expansion.
\item
The functional approach directly leads to Feynman rules on the CTP, which appear to
be more easily tractable than evaluating the commutators that occur in
the operator approach. Moreover, the CTP techniques are closely related to
the real-time formalism in equilibrium field theory, such that it is comparably
simple to obtain the relevant self-energies in finite temperature or
non-equilibrium backgrounds for leptogenesis.
\item
Related to the previous point, all self-energies relevant for leptogenesis
have been evaluated in the CTP formalism (see e.g. Ref~\cite{Garbrecht:2013urw} for the self-energy of
relativistic RH neutrinos relevant in the ARS scenario), such that this approach is
self-contained.
\end{itemize}
We emphasize however that given the the standard perturbative expansion
appropriate for leptogenesis
as a weakly coupled theory and provided the same
approximations are made,
operator-based and CTP approaches will also yield identical results.
At the present state of the art, the choice of method is therefore a matter of
preference, calculational transparency or convenience rather than accuracy.

After these more general remarks, we now apply the program sketched
in \fref{fig:fundamental-to-fluid} to derive the fluid equations
for ARS leptogenesis in the CTP framework. The present emphasis is on the evolution of the state of the
RH neutrinos, while the flavor dynamics of the active doublets
from the perspective of the Schwinger-Keldysh CTP formalism is presented in detail
in the accompanying article on flavored leptogenesis~\cite{leptogenesis:A01}. For now, we note that
the ARS scenario is a variant of \emph{purely}
flavored leptogenesis because the initially produced total $B-L$ asymmetry vanishes
while there are asymmetries in the individual flavors.
Further, we are in the \emph{fully} flavored regime since the temperatures are
low enough such that there are no coherent
correlations between the doublet flavors because these are erased immediately by the
SM lepton Yukawa couplings, which mediate reactions that are fast compared to
the Hubble rate.
The source for the asymmetry, i.e. the time derivative of the lepton charge
density is given by
\begin{align}
\label{Source:flavoured:1}
S_{\alpha\beta}=-\sum\limits_{i\not=j}\lambda_{\alpha i} \lambda^*_{\beta j} \int\frac{d^4 k}{(2\pi)^4}
{\rm tr}
\left[
P_{\rm R}{i}\delta S_{N ij}(k) 2P_{\rm L}
{\hat{\Sigma}\!\!\!/}^{\cal A}_N(k)
\right]\,,
\end{align}
where
the reduced self-energy
is defined through
\begin{align}\label{hatSigmaNDef}
{\Sigma}\!\!\!/_N&= g_w \hat{\Sigma}\!\!\!/ \left(
\lambda^T \lambda^* P_{\rm R}
+
\lambda^\dagger \lambda P_{\rm L}
\right) 
\,,
\end{align}
i.e. with the Yukawa couplings stripped,
and the spectral self-energy ${{\Sigma}\!\!\!/}^{\cal A}_N(k)$ can be interpreted as the cut part
of the Higgs-lepton loop at finite temperature~\cite{Garbrecht:2011aw}.
The indices $\alpha,\beta=e,\mu,\tau$ denote the lepton flavor, and we can delete the
off-diagonal correlations since we are in the \emph{fully} flavored regime.
Since the two scenarios are closely related,
the CTP derivation of the expression~\eqref{Source:flavoured:1} is discussed
in the accompanying review article on resonant leptogenesis~\cite{leptogenesis:A03}.

Now we turn to the main focus of the present section, the computation
of ${i}\delta S_{N\,{ij}}$, which is the out-of-equilibrium
component of the statistical propagator of the RH neutrinos $N$. 
The present discussion constitutes a summary of the elaboration
in Ref.~\cite{Drewes:2016gmt}, where more technical details are provided.
The indices $i,j$ denote
the RH neutrino flavors, while spinor indices are suppressed. Note that for the RH neutrinos,
the off-diagonal correlations are crucial because these provide the CP-even phases
that are required in order to obtain a CP-violating effect.

\paragraph{Formulation of the kinetic equations}
The Schwinger-Dyson equations on the CTP are most straightforwardly formulated in
terms of two-point functions.
Let $\Psi$ be a spinor field with mass matrix $M$, then the Wightman functions are given by
\begin{eqnarray}
{i} S^{>}_{\rho\sigma}(x_{1},x_{2})=\langle \Psi_{\rho}(x_{1})\bar{\Psi}_{\sigma}(x_{2})\rangle \ , \quad
{i} S^{<}_{\rho\sigma}(x_{1},x_{2})=-\langle \bar{\Psi}_{\sigma}(x_{2})\Psi_{\rho}(x_{1})\rangle\,,
\label{SbackA}
\end{eqnarray}  
where $\rho$ and $\sigma$ are spinor indices, which we suppress in the following. Flavor
indices can be introduced straightforwardly. From these basic correlators,
we can construct the following Green's functions:
\begin{subequations}
\begin{align}
S^{\mathcal{A}}(x_{1},x_{2})&\equiv \frac{{i}}{2}\left(S^{>}(x_{1},x_{2})-S^{<}(x_{1},x_{2})\right)\label{SMinus}\;\; \textnormal{spectral function},\\
S^{+}(x_{1},x_{2})&\equiv \frac{1}{2}\left(S^{>}(x_{1},x_{2})+S^{<}(x_{1},x_{2})\right)\label{Splus}\;\;\textnormal{statistical propagator},\\
{i} S^{R}(x_{1},x_{2})&=2\theta(t_1-t_2)S^{\mathcal{A}}(x_{1},x_{2})\;\;\textnormal{retarded propagator},\label{Sretarded}
\\
{i} S^{A}(x_{1},x_{2})&=-2\theta(t_2-t_1)S^{\mathcal{A}}(x_{1},x_{2})\;\;\textnormal{advanced propagator},\label{Sadvanced}
\\
 S^{H}(x_{1},x_{2})&=\frac12\left(S^R(x_1,x_2)+S^A(x_1,x_2)\right)\;\;\textnormal{Hermitian propagator}.\label{Shermitian}
\end{align}
\end{subequations}
Corresponding quantities can be defined for other two-point functions, in particular
for the self-energies $\slashed\Sigma$, and we identity these by the same superscripts.

We take as the starting point for the derivation of kinetic equations Schwinger-Dyson equations
on the CTP, which are exact in principle, cf. the scheme outlined in \fref{fig:fundamental-to-fluid}.
We then make the transition from position into Wigner
space through a Fourier transformation with respect to the relative coordinate
$x_1-x_2$ while keeping the average coordinate $X=(x_1+x_2)/2$. For weakly coupled particles,
the spectral function then exhibits a quasi-particle peak at the mass shell, and the Wigner space
coordinate $k$ can be interpreted as the four momentum. In Wigner space, the
convolution integrals take the form of a Moyal product~\cite{Groenewold:1946kp,Moyal:1949sk,Greiner:1998vd,Prokopec:2003pj}. The gradient expansion
consists (in the present spatially homogeneous case) of dropping temporal derivatives
that are of order of the Hubble rate (if the associated process is close to equilibrium)
or the particular relaxation rate (if the process is far from equilibrium). At leading
order, the Moyal product corresponds to an ordinary multiplication. Applying this
truncation, the Schwinger-Dyson equations on the CTP read
\begin{subequations}
\begin{align}
\Big(\slashed{p} +\frac{{i}}{2} \gamma_0\partial_t - M \Big)S^\mathcal{A}- 
\left(\slashed{\Sigma}^H S^\mathcal{A} + \slashed{\Sigma}^\mathcal{A} S^H\right)&=0\,,\label{kbe1}\\
\Big(
\slashed{p} +\frac{{i}}{2} \gamma_0\partial_t - M
\Big)S^+ - \slashed{\Sigma}^H S^+ - \slashed{\Sigma}^+ S^H &= \frac{1}{2}\left(\slashed{\Sigma}^> S^< - \slashed{\Sigma}^< S^>\right)\label{kbe2}\,,
\end{align}
\end{subequations}
where all two-point functions depend on $(k,X)$, and Eq.~\eqref{kbe2} is known as
the Kadanoff-Baym equation. We also write $t\equiv X^0$ for the time coordinate, and
derivatives with respect to $X^i$, $i=1,2,3$, do not occur provided spatial homogeneity holds.

Decomposing the Kadanoff-Baym equation~(\ref{kbe2}) into its Hermitian and anti-Hermitian
part, we obtain the constraint and kinetic equations
\begin{subequations}
\begin{align}\label{constrainedSplus0}
\{ \mathcal{H}, \mathcal{S}^+\} 
-\{\mathcal{N},\mathcal{S}^H\}
&=\frac{1}{2}\left([\mathcal{G}^>,\mathcal{S}^<]-[\mathcal{G}^<,\mathcal{S}^>]\right)
\,,\\
{i} \partial_t \mathcal{S}^+ +[\mathcal{H},\mathcal{S}^+]
-[\mathcal{N},\mathcal{S}^H]
&=\frac{1}{2}\left(\{\mathcal{G}^>,\mathcal{S}^<\}-\{\mathcal{G}^<,\mathcal{S}^>\}\right)\,,
 \label{kineticSplus0}%\ {\rm \ ( kinetic \ equation)}
\end{align}
\end{subequations}
where
\begin{eqnarray}
\nonumber
\mathcal{S}^+&\equiv&{i}\gamma^0 S^+ \ , \ \mathcal{S}^H\equiv {i}\gamma^0 S^H \ , \ \mathcal{H}\equiv(\slashed{p}-\slashed{\Sigma}^H-M)\gamma^0 %\ , \mathcal{G}\equiv\Sigma^-\gamma^0,
\,,\\ 
\mathcal{G}^>&\equiv&\slashed{\Sigma}^>\gamma^0 \  , \ \mathcal{G}^<\equiv\slashed{\Sigma}^<\gamma^0 , \ \mathcal{G}\equiv\frac{{i}}{2}(\mathcal{G}^>-\mathcal{G}^<) ,
\ \mathcal{N}\equiv \slashed{\Sigma}^+\gamma^0\,.
\end{eqnarray}
In order to isolate the non-equilibrium dynamics on top of the equilibrium background, we further decompose
\begin{equation}
\mathcal{H}=\bar{\mathcal{H}}+\delta\mathcal{H} \ , \ \mathcal{G}=\bar{\mathcal{G}} + \delta\mathcal{G}\,, \mathcal{S}^+=\bar{\mathcal{S}}^+ + \delta{\mathcal{S}}\label{splitSN}\,,
\end{equation}
where $\bar{\mathcal{H}}$ and $\bar{\mathcal{G}}$ are $\mathcal{H}$ and $\mathcal{G}$ evaluated in thermal equilibrium and with vanishing chemical potentials.
The two-point function
$\bar{\mathcal{S}}^+=(\bar{\mathcal{S}}^>+\bar{\mathcal{S}}^<)/2$ is a time-independent static solution that satisfies the algebraic equation
\begin{eqnarray}
[\bar{\mathcal{H}},\bar{\mathcal{S}}^+]
-[\bar{\mathcal{N}},\bar{\mathcal{S}}^H]
=\frac12\left(\{\bar{\mathcal{G}}^>,\bar{\mathcal{S}}^< \}-\{\bar{\mathcal{G}}^<,\bar{\mathcal{S}}^> \}\right)\,.
\end{eqnarray}
%and
%\begin{eqnarray}
%[ \mathcal{H},\bar{\mathcal{S}}^\mathcal{A} ] &=& \{ \mathcal{G} , \mathcal{S}^H \}.
%\end{eqnarray}
Eventually, to leading order in gradients, couplings and chemical potentials over
temperature, the non-equilibrium part of the statistical propagator is
a solution to the kinetic equation
\begin{align}
\nonumber
\partial_t \delta\mathcal{S} 
&=
-\partial_t \bar{\mathcal{S}}^+
+{i}[\bar{\mathcal{H}},\delta{\mathcal{S}}]
+{i}[\delta\mathcal{H},\bar{\mathcal{S}}^+]
-{i}[\delta\mathcal{N},\bar{\mathcal{S}}^H]
-\{\bar{\mathcal{G}},\delta\mathcal{S} \}\\
&-\frac{{i}}{2}\left(\{\delta\mathcal{G}^>,\bar{\mathcal{S}}^< \}-\{\delta\mathcal{G}^<,\bar{\mathcal{S}}^> \}\right)
\,.\label{generalkinetic}
\end{align}
The deviation from equilibrium due to the expansion of the Universe enters through a non-vanishing
$\partial_t \bar{\mathcal{S}}^+$,\footnote{The equilibrium distribution changes due to Hubble expansion. We emphasize however that the non-equilibrium relevant for ARS leptogenesis is due to
the vanishing initial conditions of the RH neutrinos.} and the first of the commutator terms  leads to phase oscillations
in the off-diagonal flavor correlations.

We now specialize to the case of ARS leptogenesis, where $\delta{\mathcal S}_N$ denotes
the deviation of the propagators of the RH neutrinos (and their flavor correlations)
from equilibrium. For that purpose, we can truncate the equations at second order
in the RH neutrino Yukawa couplings $\lambda$, and we expand up to linear order
in the chemical potentials of the doublet leptons and the Higgs bosons. In this
approximation, the terms involving $\delta {\mathcal H}$ and $\delta {\mathcal N}$
in Eq.~\eqref{generalkinetic} can be dropped, whereas the remaining contributions
can be wrapped up as
\begin{align}
\label{eq:kinetic_eq_tensorial}
\partial_t \delta \mathcal{S}_N =&2\frac{\partial_t f_F}{1-2f_F}\bar{\mathcal{S}}^+_N 
+{i} [\bar{\mathcal{H}}_N,\delta \mathcal{S}_N]-\{\bar{\mathcal{G}}_N,\delta \mathcal{S}_N\}
\notag\\
-&\frac{2}{1-2f_F}\sum_{a=e,\mu,\tau} \frac{\mu_{\ell a}+\mu_\phi}{T} \{\tilde{\mathcal{G}}^{a}_N, \bar{\mathcal{S}}^+_N\}\,,
\end{align}
where
\begin{eqnarray}
\tilde{\mathcal{G}}^{a}_N=-g_w f_F[1-f_F]
\hat{\Sigma}\!\!\!/^\mathcal{A}_N
\left(
\lambda^T_{i\alpha} \lambda_{\alpha j}^*P_{\rm R} - \lambda^\dagger_{i\alpha}\lambda_{\alpha j} P_{\rm L}
\right)\gamma^0\,,
\end{eqnarray}
and $f_F=f_F(k^0)=1/(e^{k^0/T}+1)$ denotes the Fermi-Dirac distribution.

\paragraph{Separation of the spinor structure}
To this end, $\delta \mathcal{S}_N$ is still a product of a spectral function
with a statistical distribution.
First, the spinor structure can be factorized apart as
\begin{eqnarray}
\label{eq:tensor_dec_propagator}
-{i} \gamma^0 \delta S_N = \sum_h \frac{1}{2} P_h \left(g_{0h} + \gamma^0 g_{1h} - {i} \gamma^0\gamma^5 g_{2h} -\gamma^5 g_{3h}\right)\,,
\end{eqnarray}
with the helicity projectors 
\begin{equation}
P_h \equiv \frac{1}{2}\left(1+h\hat{\bf k}\gamma^0\pmb{\gamma}\gamma^5\right)\,,
\end{equation}
and where the functions $g$ are Hermitian matrices in the space of RH neutrino flavors.
In this decomposition, there occur also left-handed components in the correlations,
i.e. $\delta S_N$ can be constructed from Majorana spinors. Due to the chiral nature
of the Yukawa interactions, these components are however projected out, such that
it would be also possible, but notationally perhaps more cumbersome, to use purely RH spinors

We can now use the constraint equation \eqref{constrainedSplus0} in order to 
establish to leading order in $M_N/k_0$ and $\lambda$ the relations
\begin{align}
\label{eq:constraint_propagator}
g_{1h}&=\frac{1}{2k^0}\left( \{\Re M_N,g_{0h}\} +
[{i} \Im M_N,g_{3h}] \right)\,, \\
\notag
g_{2h}&=\frac{1}{2{i} k^0}\left([\Re M_N,g_{3h}]+\{{i} \Im M_N,g_{0h}\} \right)\,,\\
\notag g_{3h}&=h\mathrm{sign}(k^0)g_{0h}\,.%\label{LorentzComponentRelations}
\end{align}
Now, the kinetic equations can be written in terms of ${g}_{0h}$ and $\bar{g}_{0h}$ only,
a simplification since the spinor structures have been dropped:
\begin{align}
\label{eq:off_shell}
\partial_t g_{0h} =& 2\frac{\partial_t f_F}{1-2f_F} \bar{g}_{0h} -\frac{{i}}{2} [ {\rm H}_N,g_{0h}] 
-
\frac12 \{ \Upgamma_N,g_{0h}
\} \notag\\
-&\frac12 \frac{2}{1-2f_F}\sum_{a=e,\mu,\tau}
\frac{\mu_{\ell a}+\mu_\phi}{T}\{ \tilde{\Upgamma}_N^a,\bar{g}_{0h}
\}\,,
\end{align}
where
\begin{subequations}
\begin{align}
{\rm H}_N&=2 g_w \left(
{\rm Re}[\lambda^T \lambda^*]\frac{k\cdot\hat{\Sigma}_N^{H}}{k^0}
-{i}h {\rm sign}(k^0){\rm Im}[Y^*Y^t]\frac{k\cdot \hat \Sigma_N^{H}}{k^0}
\right)
\\&+\frac{1}{k^0}\left( \Re[M_N^\dagger M_N] +
{i} h \mathrm{sign}(k_0) \Im[M_N^\dagger M_N] \right)\,,
\\
\Upgamma_N%&=& (1-h\frac{k^0}{\textbf{k}})\Sigma^\mathcal{A}_L + (1+h\frac{k^0}{\textbf{k}})\Sigma^\mathcal{A}_R \\
 &=2 g_w \left(
{\rm Re}[\lambda^T \lambda^*]\frac{k\cdot \hat \Sigma_N^{\mathcal{A}}}{k^0}
-{i}h {\rm sign}(k^0){\rm Im}[Y^*Y^t]\frac{k\cdot \hat \Sigma_N^{\mathcal{A}}}{k^0}
\right) \, ,\\
(\tilde{\Upgamma}_N^a)_{ij} &=  2 h
g_w f_F(1-f_F) \left(
{\rm sign}(k^0){\rm Re}[\lambda^T_{i\alpha}\lambda^*_{\alpha j}]\frac{ k\cdot \hat \Sigma_N^{\mathcal{A}}}{k^0}-{i} h {\rm Im}[\lambda^T_{i\alpha}\lambda^*_{\alpha j}]\frac{k\cdot \hat \Sigma_N^{\mathcal{A}}}{k^0}
\right)\,.
\end{align}
\end{subequations}

\paragraph{Quasi-particle approximation}
Eventually, we have reached the point where we can factorize the spectral function from
the quasiparticle distributions. For that purpose, we can once more take Eq.~\eqref{constrainedSplus0} and establish that
\begin{eqnarray}
\label{eq:quasiparticle_app}
\bar{g}_{0h}(k)_{ij}\approx
-\frac{1-2f_F}{2} \delta_{ij} 
2\pi\delta(k_0^2-\Omega_i^2)2k^0\mathrm{sign}(k^0)\,
\end{eqnarray}
and
\begin{align}
\label{eq:quasiparticle_app:2}
g_{0h}=-\frac{2}{1-2f_F}\bar{g}_{0h}\delta f_{0h}\,.
\end{align}
We emphasize that this quasi-particle approximation is derived from the
constraint equation and does not correspond to an ad hoc ansatz.
The mass-square term $\Omega^2_i$ can be derived from the constraint equations or alternatively
from the equations for the retarded and advanced propagators. It includes the tree-level Majorana
masses as well as thermal corrections. However, since we work in the relativistic regime where
$T\gg M_{Ni}$, we can approximate $\Omega_i\approx 0$ up to a small error due to kinematic corrections.

We can interpret $\delta f_{0h}(|\mathbf k|,\textbf{k}) + f_F(|\mathbf k|)$ as the distribution function
of particles and 
$1 - \delta f_{0h}(-|\mathbf k|,\textbf{k}) - f_F(-|\mathbf k|)$ of antiparticles.
For the heavy neutrinos, the Majorana condition in the mass eigenbasis implies 
\begin{align}
\label{Majorana:dist}
\delta f_{0h}(-k^0)=\delta f_{0h}^*(k^0)\,,
\end{align}
such that we only need to track the particle distribution, i.e. we can
restrict to the case $\mathrm{sign}(k^0)=1$.

Taking the zeroth moment
\begin{align}
\int  \!\frac{\mathrm{d} k^0}{2\pi} g_{bh}=\delta f_{bh}\,,
\end{align}
we obtain a kinetic equation for distribution functions,
\begin{align}
\partial_t \delta f_{0h} &= - \partial_t f_F -\frac{{i}}{2} [ {\rm H}_N,\delta f_{0h}] 
-
\frac12 \{ \Upgamma_N,\delta f_{0h}
\}+\sum_{a=e,\mu,\tau}
\frac{\mu_{\ell a}+\mu_\phi}{T}\tilde{\Upgamma}_N^a\,.
\end{align}
Here, all quantities are evaluated on the zero-mass shell $k^0=|\mathbf k|$, as it is valid in
the relativistic approximation.

The expansion of the Universe can be accounted for by using conformal time $\eta$, which
can be normalized during the radiation era as $\eta=1/T$. It then follows that
\begin{align}
\delta f_{0hij}^\prime&
+\frac{{i}}{2} \left[{\rm H}_N,\delta f_{0h} \right]_{ij}
+(f^{\rm eq})^{\prime}_{ij}
= -\frac12\left\{\Upgamma_N, \delta f_{0h}\right\}_{ij} +\sum_{a=e,\mu,\tau}\frac{\mu_{\ell a}+\mu_\phi}{T}(\tilde{\Upgamma}_N^a)_{ij}\,,
\label{evolution:sterile}
\end{align}
where a prime denotes a derivative with respect to $\eta$. This equation
for the distribution function is the analogue of Eq.~\eqref{QB:Valencia} derived
in the operator formalism.
The momentum $|\mathbf k|$ now denotes a comoving momentum from which
we obtain the physical one when dividing by the scale factor $a=a_{\rm R}\eta$, where
$a_{\rm R}$ is an arbitrary parameter during radiation domination. Accordingly, $T$ is a comoving
temperature, and we need to divide by $a$ to obtain the physical temperature. Equation~\eqref{evolution:sterile} is a Boltzmann-like or quantum-Boltzmann equation that corresponds to the second step
shown in \fref{fig:fundamental-to-fluid}. It can also be useful to trade $\eta$ for the
proportional but dimensionless parameter $z=T_{\rm ref}/T=\eta T_{\rm ref}$. For the
present problem, we choose $T_{\rm ref}=130\,{\rm GeV}$, which coincides with
sphaleron freeze-out.

Since $a_{\rm R}/a=T$ can be interpreted as a comoving temperature, the equilibrium distribution for massless sterile neutrinos that appears in Eq.~(\ref{evolution:sterile}) is given by
\begin{align}
f^{\rm eq}=\frac{1}{\mathrm{e}^{|\textbf{k}|/a_{\rm R}}+1}\,,
\end{align}
which is independent of conformal time, such that the term $f^{\rm eq '}$ in Eq.~(\ref{evolution:sterile})
vanishes.
This reflects a key difference between the ARS scenario and standard leptogenesis in the strong washout regime:
while in the latter case, the relevant deviation from equilibrium is due to the fact that red shift
drives the distribution of massive RH neutrinos away from the Fermi-Dirac form, in the present case, the
deviation from equilibrium is given as an initial condition, where it is usually assumed that the
RH neutrino distributions are vanishing in first place.

\paragraph{Reduction to fluid equations}
Equation~(\ref{evolution:sterile}) would render leading order accurate predictions,
given the approximations that we have discussed to this end as well as the required input
from the loop functions. The reduced Hermitian self energy $\hat\Sigma^{H \mu}$ is simply given by the well known real
part of the thermal loop made up from a Higgs boson and a doublet lepton~\cite{Weldon:1982bn} (cf. the discussion in \sref{sec:dispersive}), whereas
the reduced spectral self energy  $\hat\Sigma^{{\cal A} \mu}$ is dominated by a
cut that involves gauge radiation or top-quark pair-production~\cite{Anisimov:2010gy,Besak:2012qm,Garbrecht:2013urw} (cf. the discussion in \sref{sec:collisionrates}). Nonetheless, a numerical solution
would be costly because the individual momentum modes would have to be tracked and moreover,
these modes couple among one another
via the backreaction of the doublet lepton asymmetries that affect the
helicity asymmetries of the RH neutrinos.

Therefore, we derive from Eq.~(\ref{evolution:sterile}) a fluid equation for the number densities of the RH neutrinos by taking the zeroth moment over the
momentum $\mathbf{k}$. It can be
formulated in terms of
the equilibrium number density
\begin{align}
\label{n_eq}
n^{\rm eq}=\int \!\frac{\mathrm{d} ^3k}{(2\pi)^3}f^{\rm eq}=\frac{3}{4\pi^2}a_{\rm R}^3\zeta(3)
\end{align}
and the deviations from equilibrium
\begin{align}
\delta n_{hij}=\int \!\frac{\mathrm{d} ^3k}{(2\pi)^3}\,\delta  f_{0hij}(\textbf{k})\,.\label{nDeviationDef}
\end{align}
Since the RH neutrinos are relativistic, rather than taking higher moments,
a suitable procedure for going beyond the zeroth moment approximation
would be to solve the Boltzmann equations for a grid of momentum modes,
making use of spatial isotropy. For the reasons described above, phenomenological
studies to this end rely on the zeroth-moment approximation, and one should
keep in mind that this amounts to a theoretical uncertainty of order one.

In particular, this uncertainty is due to the fact that when integrating
Eq.~\eqref{evolution:sterile} over $d^3 k$, dependencies on $|\mathbf{k}|$
do not only enter via $\delta f_{0h}$ and $f^{\rm eq}$ but also
through ${\rm H}_N$, $\Upgamma_N$ and $\tilde\Upgamma_N$,
such that extra factors of $1/|\textbf{k}|$ occur. These originate,
e.g., from the contributions
\begin{align}
\nonumber
\frac{k\cdot\hat\Sigma^{H}_N}{|\mathbf{k}|}=\frac{a_{\rm R}^2}{16 |\mathbf{k}|}\,.
\end{align}
To proceed with such terms, we replace $1/|\textbf{k}|$ with its average value
\begin{align}
\label{k:av}
\left\langle\frac{1}{|\mathbf{k}|}\right\rangle\equiv \frac{1}{{n^{\rm eq}}}\int \!\frac{\mathrm{d} ^3k}{(2\pi)^3}\frac{1}{|\mathbf{k}|}f^{{\rm eq}}(\mathbf k)=\frac{\pi^2}{18 a_{\rm R}\zeta(3)}\,.
\end{align}
For the dissipative terms, we make the replacement
\begin{align}
\label{col:av}
\frac{k\cdot\hat\Sigma^{\cal A}_N}{|\textbf{k}|}\to \frac{\gamma_{\rm av} a_{\rm R}}{2 g_w}\,,
\end{align}
with the averaged rate $\gamma_{\rm av}\equiv\Gamma_{\rm av}/T$, which turns out
to be dominated by the $t$-channel exchange of charged leptons associated with
gauge radiation. This rate has been estimated or computed using a variety of methods in Refs.~\cite{Anisimov:2010gy,Kiessig:2010pr,Laine:2011pq,Fidler:2011yq,Salvio:2011sf,Biondini:2013xua,Besak:2012qm,Bodeker:2014hqa,Laine:2013lka,Garbrecht:2013urw,Ghisoiu:2014ena,Ghiglieri:2016xye}. We emphasize that it is possible to derive this rate using
CTP methods~\cite{Garbrecht:2013urw}, such that we can comply with the
plan to remain with our derivations inside the framework of the CTP formalism.
In Ref.~\cite{Garbrecht:2014bfa}, the value $\gamma_{\rm av}=0.012$ is used, which is
based on the derivations of Refs.~\cite{Besak:2012qm,Garbrecht:2013urw}.

For an electroweak crossover and a large degree of degeneracy in the eigenvalues of the Majorana mass matrix $M_N$, the evolution of the vacuum expectation value $v(z)$ of the Higgs field
may become relevant around electroweak temperatures. When inserting the vacuum expectation values perturbatively and
neglecting the small admixture of doublet leptons to the sterile neutrinos for $v(z)\not=0$,
the Hermitian part of the self-energy can be decomposed as follows
\begin{align}
\nonumber
\frac{k\cdot\hat\Sigma^{H}_N}{|\textbf{k}|}+
\frac{a^2 v^2(T)}{g_w |\textbf{k}|}=\frac{a_{\rm R}^2}{16 |\textbf{k}|}+
\frac{a^2 v^2(T)}{g_w |\textbf{k}|} \to \frac{a_{\rm R}}{2 g_w} 
\left(
\mathfrak{h}_{\rm th}+\mathfrak{h}_{\rm EV}(z)
\right)\,,
\end{align}
where, when using the procedure implied by Eq.~(\ref{k:av}), it follows
\begin{align}
\mathfrak{h}_{\rm EV}(z)=\frac{2\pi^2}{18\zeta(3)}\frac{v^2(z)}{T_{\rm ref}^2}z^2 \ , \ \mathfrak{h}_{\rm th}\approx 0.23\,.
\end{align}

Altogether, the integration  of Eq.~(\ref{evolution:sterile}) over the three momentum $\mathbf{k}$, leads to a fluid equation for the sterile number densities
\begin{align}
\label{diff:sterile}
\frac{\mathrm{d}}{\mathrm{d} z}\delta n_{h} = -\frac{{i}}{2}[H^{\rm th}_N+z^2 H^{\rm vac}_N,\delta n_{h}]-\frac{1}{2}\{\Gamma_N,\delta n_{h}\}+\sum_{a=e,\mu,\tau}\tilde{\Gamma}^a_N \left(q_{\ell a}+\frac12 q_{\phi}\right)\,,
\end{align}
where the averaged rates appearing here are
\begin{subequations}
\begin{align}
%\label{RHN:rates}
H^{\rm vac}_N &=
\frac{\pi^2 }{18 \zeta(3)}\frac{a_{\rm R}}{T_{\rm ref}^3}
\left(\Re[M_N^\dagger M_N] + {i} h  \Im[M_N^\dagger M_N]\right)\,,\nonumber\\
H^{\rm th}_N&=[\mathfrak{h}_{\rm th}+\mathfrak{h}_{\rm EV}(z)]\frac{a_{\rm R}}{{T_{\rm ref}}}
\left(\Re[\lambda^T \lambda^*]-{i} h\Im [\lambda^T \lambda^*]\right)\,,\nonumber\\
%\label{avg:decayrate}
\Gamma_N &=\gamma_{\rm av} \frac{a_{\rm R}}{{T_{\rm ref}}}
\left(\Re[\lambda^T \lambda^*]-{i} h \Im [\lambda^T\lambda^*]\right)\,,\nonumber\\
%\label{avg:backreaction}
(\tilde{\Gamma}^a_N)_{ij}&= \frac{h}{2}\gamma_{\rm av} \frac{a_{\rm R}}{T_{\rm ref}}
\left(
\Re [\lambda^T_{i\alpha}\lambda^*_{\alpha j}] - {i} h  \Im [\lambda^T_{i\alpha}\lambda^*_{\alpha j}]
\right)\,.\nonumber
\end{align}
\end{subequations}
The result~\eqref{diff:sterile} 
corresponds to the final box shown in \fref{fig:fundamental-to-fluid} as
well as the momentum-averaged Eqs.~\eqref{eq:rhonrhonbarav} found
in the operator-based approach.

Coming back to the order one inaccuracy in these equations, we note that it is essentially
due to the fact that the non-equilibrium distribution $\delta f_{0h}(\mathbf k)$ of the RH neutrinos is not known unless
the momentum modes are tracked separately. The remaining particles such as charged leptons,
Higgs and gauge bosons that take part in the relevant interactions are maintained in kinetic
equilibrium by the gauge interactions. Therefore their distributions are known and the momentum integrals can be evaluated accurately. Regarding the RH neutrino distributions,
the situation in the ARS case is also different from standard leptogenesis in the strong
washout regime because in the latter case, the momentum can be assumed to be much
smaller than the temperature due to the Maxwell suppression. The momentum integral
over these non-relativistic RH neutrinos can therefore be evaluated without incurring an
order one inaccuracy.

\paragraph{Leptonic charges}
Another decisive difference between strong washout and ARS leptogenesis lies within the
fact that in the former case, the Majorana mass of the RH neutrinos close-to-maximally
violates the chirality and hence the fermion number, whereas for the relativistic RH neutrinos
of the ARS case, this is a sub-dominant effect. Fermion number is therefore
conserved to leading approximation, and it is therefore useful to 
define the even and odd parts of the number densities
\begin{align}
n_{ij}^{{\rm even},{\rm odd}}=\frac12\left( \delta n_{+ij}\pm \delta n_{-ij}\right)
\end{align}
and the RH neutrino charge
\begin{align}\label{qNDef}
q_{Nij}=\delta n_{+ij}-\delta n_{-ij}
=2n_{ij}^{\rm odd}
\,.
\end{align}
This charge is produced due to the backreaction from the asymmetry in doublet leptons
and in turn also appears as a source for the doublet asymmetry.
Including also the CP-violating source~\eqref{Source:flavoured:1} and the washout,
the evolution of the SM-conserved charge densities $\Delta_\alpha=B/3-L_{\alpha}$ is given by
\begin{align}
\label{evolution:active}
\frac{{\rm d} \Delta_{\alpha}}{{\rm d} z}=-\frac{a_{\rm R}}{T_{\rm ref}}W_\alpha \left(q_{\ell\,\alpha}+\frac12 q_\phi-q_{Nii} \right)+
\frac{1}{T_{\rm ref}}S_{\alpha}\,,
\end{align}
where the CP-violating source follows from the flavor-diagonal part of Eq.~\eqref{Source:flavoured:1} (because we are in the fully flavored regime),
\begin{align}
\label{Source}
S_\alpha=2\frac{\gamma_{\rm av}}{g_w} a_{\rm R}\sum\limits_{\overset{i,j}{i\not=j}}\lambda_{\alpha i}\lambda^\dagger_{j\alpha}
\left[{i}{\rm Im}(\delta n^{\rm even}_{ij})+{\rm Re}(\delta n^{\rm odd}_{ij})\right]\,,
\end{align}
and the washout rate (that is complementary to the damping rates for sterile neutrinos $\Gamma_N$) is given by
\begin{align}
W_\alpha&=\frac{\gamma_{\rm av}}{g_w}\sum_i \lambda_{\alpha i}\lambda^\dagger_{i\alpha}\,.
\end{align}
In Eq.~\eqref{evolution:active}, we also
take account of fast spectator processes
that redistribute the asymmetries among the SM degrees of freedom according
to
\begin{align}
\label{linrel:spectators:2}
q_{\ell\,\alpha}=-\sum\limits_\beta C_{\alpha\beta}\Delta_\beta\,,\quad
q_\phi=\sum\limits_\alpha B_\alpha \Delta_\alpha\,,\quad
B=-\frac{8}{79}\left(
\begin{array}{ccc}
1&1&1
\end{array}
\right)
\,,
\end{align}
where the matrix $A$ is given in Eq.~\eqref{linrel:spectators}.

\section{Analytical expansions}
\label{section:analytical}

The Eqs.~(\ref{eq:rhonrhonbarav}) are challenging to treat numerically because they involve two or more relevant time scales, in particular $t_{\rm osc}(ij)$ and $t_{\rm eq}(\alpha)$ that can 
be widely different. For this reason it is useful to have some analytical understanding on the expected solutions. Several analytical approaches have been 
considered in the literature. The most straightforward one is a perturbative expansion in the Yukawa couplings, which is also a weak-washout expansion, since it 
implies that the equilibration rates, ${\mathcal O}(\lambda^2)$ (or $t_{\rm eq}(\alpha)^{-1}$) are assumed small compared to any other scale in the problem, in particular small compared to $t_{\rm osc}(ij)^{-1}$ and to $t_{EW}^{-1}$.  This implies that this approximation would fail for regions of parameter space where any flavor equilibrates before the EW phase transition $t_{\rm eq}(\alpha) < t_{EW}$. 

\subsection{Weak-washout expansion}

This approximation was first used in the original proposal of the ARS mechanism\cite{Akhmedov:1998qx} where it was found that 
\begin{eqnarray}
Y_{\Delta B} \propto 7\cdot 10^{-4}  {\Gamma\left({1\over 3}\right)^{1/3} \over 96} \gamma_N  {\rm Im}[W_{12}^* W_{13} W_{23}^* W_{22
}] 
{(y_1^2 -y_2^2)(y_1^2 -y_3^2)(y_2^2 -y_3^2) {M_P^*}^2 \over (\Delta M_{N 13}^2)^{1/3} (\Delta M_{N 12}^2)^{2/3}},\nonumber\\
\label{eq:ars}
\end{eqnarray}
where $\gamma_N \simeq \langle \gamma_N^{(0)}\rangle/T \simeq {\cal O}(10^{-3})$.
 
For $n_R=2$, Asaka and Shaposhnikov \cite{Asaka:2005pn} found in a Yukawa expansion
\begin{eqnarray}
Y_{\Delta B} \propto  7\cdot 10^{-4} {16 \pi^{3/2}  \over 
 3^{{4\over 3}}\Gamma\left({5 \over 6}\right)} \gamma_N^3\!  \sum_{\alpha,i< j}\!\! (\lambda\lambda^\dagger)_{\alpha\alpha} 
{\rm Im}\left[ \lambda_{\alpha i} (\lambda^\dagger \lambda)_{ij} \lambda^\dagger_{j\alpha}\right] {(M_P^*)^{7/3}
\over (\Delta M^2_{N ij})^{2/3} T_{EW}}.
\label{eq:as}
\end{eqnarray}
The results~\eref{eq:ars} and~\eref{eq:as} are expressed for $M_{1}<M_{2}<\cdots$.
In Ref.~\cite{Drewes:2012ma}, it was estimated that in the case when one of the flavors would be very 
far from thermalization at $T_{EW}$, while the others would have thermalized, a good approximation to the total
baryon asymmetry should be the asymmetry in the {\it slow} flavor $\alpha$:
\begin{eqnarray}
\label{asymmetry:Valencia:perturbative}
Y_{\Delta B} \propto -1.2 \cdot 10^{-4} \cdot 
{6 \over 79}~ \gamma_N^2\!\sum_{i\neq j} {\rm Im}\left[\lambda_{\alpha i} (\lambda^\dagger \lambda)_{ij} \lambda^\dagger_{j\alpha}\right]{\rm sign}(M_{j}-M_{i}) {{(M_{\rm Pl})}^{{4\over 3}}\over |\Delta M^2_{N ij}|^{{2\over 3}}}.
\label{eq:dg}
\end{eqnarray}

These three approximations give quite different results and do not even agree in the parametric Yukawa and mass dependences.

\subsection{Perturbative expansion in mixings }
\label{sec:expansion:mixings}

In Ref.~\cite{Hernandez:2015wna}, an approximation that would allow to reach the equilibration regime $t\simeq t_{\rm eq}(\alpha)$ systematically was considered. It involves an expansion in the mixing angles of the two unitary matrices that diagonalize the neutrino Yukawa coupling, $V, W$.
Since at least one state is not expected to reach equilibrium before $t_{EW}$,  we assume is the third state and take $y_3 \simeq 0$.  

The kinetic equations could be solved analytically (neglecting the running of the couplings and non-linear terms) also in the equilibration regime of the two {\it fast} flavors, $1,2$, and allowed the identification of the flavor and rephasing invariant structures expected on general grounds, as explained in \sref{sec:cpinvariants}. This allowed to clarify the relation between the original ARS result, which relies on a flavor invariant that necessary involves three
singlet families, and those found for example in Ref.~\cite{Asaka:2005pn}, which corresponds to a different flavor invariant that involves only two singlet families. In the most general case, all these contributions are present simultaneously.  

For $n_R=3$, the result is of the form
\begin{eqnarray}
Y_{\Delta B} \simeq 3\cdot 10^{-3} \sum_{CP} I_{CP}^{(i)} A_{i}(t_{EW}),
\end{eqnarray}
where $ I_{CP}^{(i)}=\{I_1^{(2)}, I_1^{(3)}, I_2^{(3)}, J_W \}$ are the following rephasing invariants:
\begin{eqnarray}
I_1^{(2)}& =&  -{\rm Im} [W^*_{1 2} V_{1 1} V^*_{2 1}  W_{2 2 } ] ,
\\
I_1^{(3)} &=&   {\rm Im} [W^*_{12} V_{13} V^*_{2 3}  W_{2 2 } ] ,
\\
I_2^{(3)} &=&  {\rm Im} [W^*_{13} V_{12} V^*_{2 2}  W_{2 3 } ],
\nonumber\\
J_W  &=& - {\rm Im}[W^*_{23} W_{22} W^*_{32} W_{33} ].
\label{eq:cpinvs}
\end{eqnarray}
Note that only four invariants appear, because two phases are actually unphysical if terms of ${\mathcal  O}(M/T)$ are neglected. 
Only the first two invariants  exist in the case $n_R=2$. The functions $A_i(t)$ are complicated functions of the Yukawas and the mass differences (for the full expressions see \cite{Hernandez:2015wna}). Only in the 
weak washout regime the dependence on the Yukawas is polynomial as in the estimates of the previous section. On the other hand simplifications arise also in the strong washout regime for the {\it fast} modes, $y_{1,2}^2 M_P^* t \gg 1$. In this limit, the leading dependence on the Yukawas  is ${\mathcal O}(y^4)$ as opposed to ${\mathcal O}(y^6)$. 

We can illustrate this with the simplest contribution from the first two invariants $I_1^{(2,3)}$. Defining $\Delta_{ij} \equiv \Delta M^2_{N ij} M_P^*/4$, neglecting spectator effects and assuming Boltzmann statistics for simplicity, at leading order in $t_{\rm osc}/t_{\rm eq}$ we get
\begin{eqnarray}
  A_{I_1^{(2)} } (t)&=&  {1 \over 4}  y_1 y_2 (y_2^2- y_1^2) \gamma_N^2  {M_P^*}^2 G_1(t),\nonumber\\
  A_{I_1^{(3)} } (t)&=& -{1 \over 4} y_1 y_2 (y_2^2- y_1^2)  \gamma_N^2 {M_P^*}^2  G_2(t),\nonumber\\
  \label{eq:AI1}
  \end{eqnarray}
with
 \begin{eqnarray}
G_1(t) &\equiv&   \left(e^{-\bar{\gamma}_2  t} - e^{-\bar{\gamma}_1 t }\right) {\rm Re}\left[i J_{20}(\Delta_{12},-\Delta_{12}, t)\right]+ ...
\label{eq:g1}
\end{eqnarray}
and
\begin{eqnarray}
G_2(t) =\left. G_1(t)\right|_{\bar{\gamma}_1=0}, 
\label{eq:g2}
\end{eqnarray}
where $\bar{\gamma}_i \equiv {4 \over 3} y_i^2 \gamma_N M_P^*$ and 
\begin{eqnarray} 
J_{20}(\alpha,-\alpha, t) \equiv \int_0^t d x_1~ e^{i {\alpha x_1^3\over 3}} \int_0^{x_1}~ d x_2 ~e^{-i {\alpha x_2^3\over 3}}.
\end{eqnarray}
The asymptotic behavior of this integral for $t \gg t_{\rm osc}$ is
\begin{eqnarray}
{\rm Im}[J_{20}(\alpha,-\alpha, \infty)] = -2 \left({2 \over 3}\right
)^{
1/3} {\pi^{3/2}\over \Gamma[-1/6]} {{\rm sign}(\alpha)\over |\alpha|^{2/3}}.
\end{eqnarray}
In the weak washout regime $\bar{\gamma}_i t \ll 1$ we find therefore
\begin{eqnarray}
\label{YBWW}
Y_{\Delta B} &\simeq&  3 \cdot 10^{-3}  \cdot {1 \over 6} \left({4 \over 3}  \right)^{4/3} {\pi^{3/2}\over \Gamma[5
/6]} \gamma_N^3  y_1 y_2 (y_2^2- y_1^2) \left((y_2^2-y_1^2) I_1^{(2)} - y_2^2 I_1^{(3)}\right) \nonumber\\\ &\times&{{M_P^*}^{7\over 3}\over (\Delta M^2_{N 12})^{2\over 3} T_{EW}} \nonumber\\\
&\simeq& 3.6\cdot 10^{-3} \gamma_N^3 \sum_{\alpha,i<j} (\lambda \lambda^\dagger)_{\alpha\alpha} {\rm Im}[\lambda_{\alpha i} (\lambda^\dagger \lambda)_{ij} \lambda^\dagger_{j\alpha} ] {{M_P^*}^{7\over 3}\over (\Delta M^2_{N ij})^{2\over 3} T_{EW}},
\end{eqnarray}
reproducing up to a numerical factor of $\sim 1/4$ the result of Eq.~\eqref{eq:as}. 

In the strong washout limit $\bar{\gamma}_i t \gg 1$ we get instead:
\begin{eqnarray}
\label{YBSW}
Y_{\Delta B} &\simeq &  -3 \cdot 10^{-3}  {1 \over 2^{1/3} 3^{4/3}} {\pi^{3/2}\over \Gamma[5
/6]} \gamma_N^2  y_1 y_2 (y_2^2- y_1^2)  I_1^{(3)} {{M_P^*}^{4\over 3}\over (\Delta M^2_{N 12})^{2\over 3} } \nonumber\\
&\simeq & -2.7 \cdot 10^{-3} \gamma_N^2  \sum_{i <j}  {\rm Im}[\lambda_{3 i} (\lambda^\dagger \lambda)_{ij} \lambda^\dagger_{j3} ] {{M_P^*}^{4\over 3}\over (\Delta M^2_{N ij})^{2\over 3}} ,
\end{eqnarray}
reproducing, up to a numerical factor of $\sim 1/8$, the result of Eq.~\eqref{eq:dg}.
The results~\eqref{YBWW} and~\eqref{YBSW} are expressed assuming $M_{N1}<M_{N2}<\cdots$.

Note  that for $n_R=3$ there are contributions from the other two invariants. In fact the contribution from the $J_W$ invariant reproduces \cite{Hernandez:2015wna} the approximate ARS result of Eq.~\eqref{eq:ars}. This contribution includes only CP phases orthogonal to those in the PMNS matrix. 

Figure~\ref{fig:companal}  shows the analytical result compared with the numerical solution to the equation for sufficiently small mixing angles.
 \begin{figure}[t]
 \begin{center}
\includegraphics[scale=0.4
]{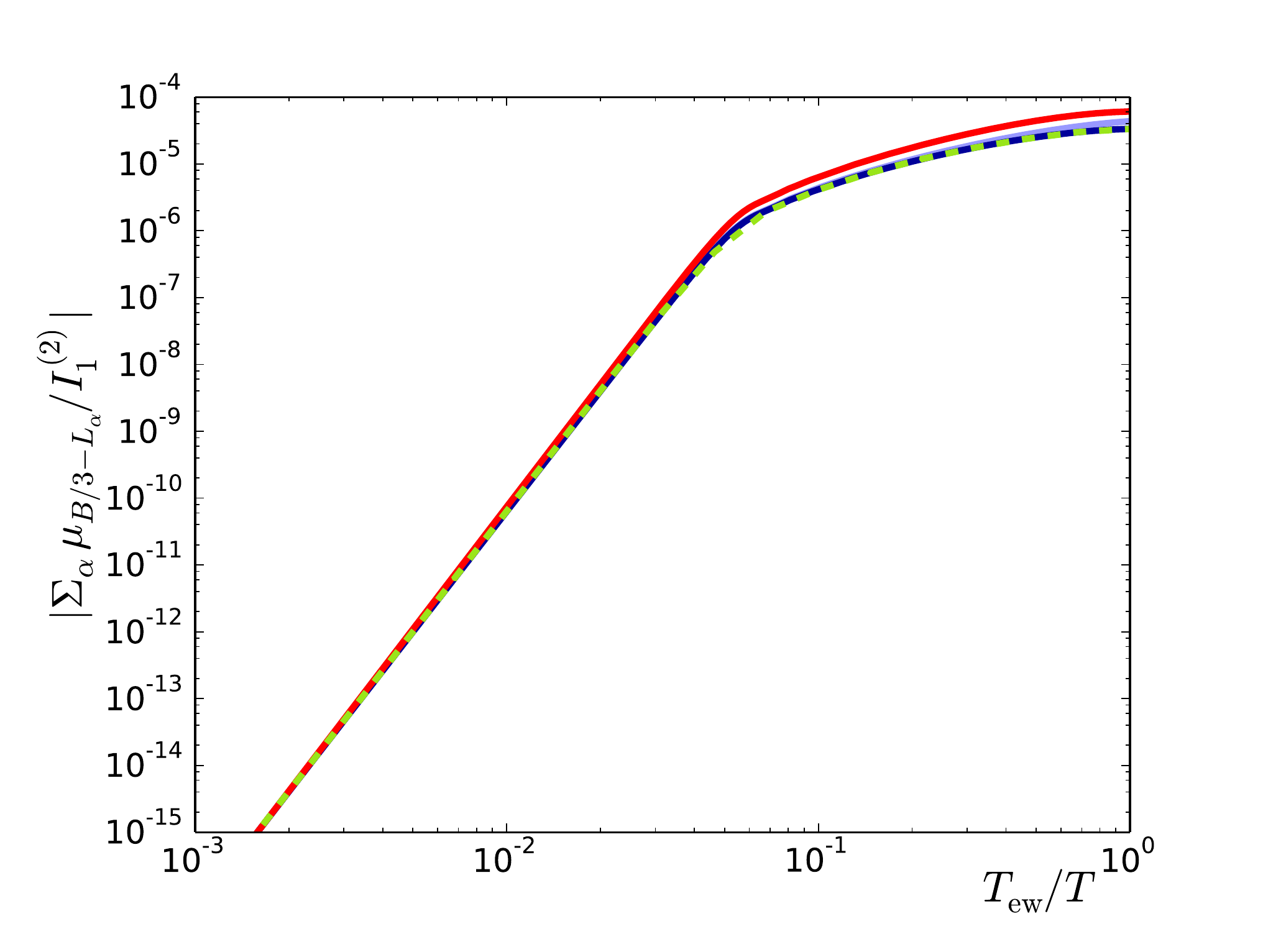} 
\caption{\label{fig:companal}  Assuming parameters for which only the simplest invariant, $I_1^{(2)}$ is non-vanishing, comparison of  the perturbative result  (dashed), the full numerical one (red), the numerical one neglecting the running of the couplings (blue) and that neglecting also non-linear terms (green).  The parameters have been chosen as $M_{1}= 1$ GeV, $M_{2} -M_{1} = 10^{-3}$ GeV, and $(y_1,y_2) = (10^{-7}, \sqrt{2} \times 10^{-7})$, and sufficiently small $V, W$ mixings. Plot taken from Ref.~\cite{Hernandez:2016kel}. }
\end{center}
\end{figure}

It is important to stress that within this perturbative expansion, where $y_3 =0$, the thermalization of the third flavor will not be reached, even though it is expected to thermalize with a rate $\propto {\rm Max}\{y_i^2 \theta_{i3}^2,i=1,2\}$, where $\theta_{ij}^2$ are the mixing angles of the matrices $V, W$. Obviously the perturbative expansion in these mixings implicitly assumes that this scale will be small compared to $t^{-1}$. This is why the limit 
$t \rightarrow \infty$ is not zero.

\subsection{The oscillatory and the overdamped regime}
\label{sec:oscioverdamped}

While it is a straightforward matter to solve the system of fluid equations for
the RH neutrino number densities as well as for $\Delta_\alpha=B/3-L_\alpha$, analytic approximations
give valuable insights, and they can prove useful in order to efficiently chart the
viable parameter space. We discuss two distinct regimes that rely on well-motivated parametric
configurations, and we refer to these as \emph{oscillatory} and \emph{overdamped}~\cite{Drewes:2016gmt}.
While the overdamped regime covers a part of the parameter space that is complementary to
the one addressed in \sref{sec:expansion:mixings}, there is an overlap for the oscillatory regime.

In order to qualify the dynamics that separates the two regimes, we note that
the first oscillation among a pair of RH neutrinos $i,j$ occurs at the time
\begin{align}
\label{time_osc}
z_\text{osc} \approx \left(a_{\rm R} |M_{i}^2 - M_{j}^2|\right)^{-1/3}T_{\rm ref}\,,
\end{align}
such that $z_{\rm osc}^3 ||H_N^{\rm vac}||={\cal O}(1)$, where
$||\cdot||$ yields the modulus of the largest eigenvalue of a matrix. This is to be compared
with the damping rate at which the RH neutrinos approach equilibrium, which
corresponds to the time scale
\begin{align}
\label{time_eq}
z_{\rm eq}\simeq T_{\rm ref}/(\gamma_{\rm av} a_{\rm R}||\lambda^T \lambda^*||)\,.
\end{align}
In analogy with a mechanical system, $z_\text{osc}\ll z_{\rm eq}$ corresponds to
the oscillatory and $z_\text{eq}\ll z_{\rm osc}$ to the overdamped regime.
Of course, in systems with more than two RH neutrinos, there are in general several
distinct oscillation frequencies and damping rates, such that for the present discussion of analytic approximations,
we mainly confine ourselves to the simple but relevant case of $n_R=2$ RH neutrinos.
The different time-scales in the oscillatory and overdamped regimes are made
visible in \fref{fig:oscillatory:overdamped}, and we next explain the analytic
approximations pertinent to these two cases.

\begin{figure}[!ht]
	\includegraphics[width=0.495\textwidth]{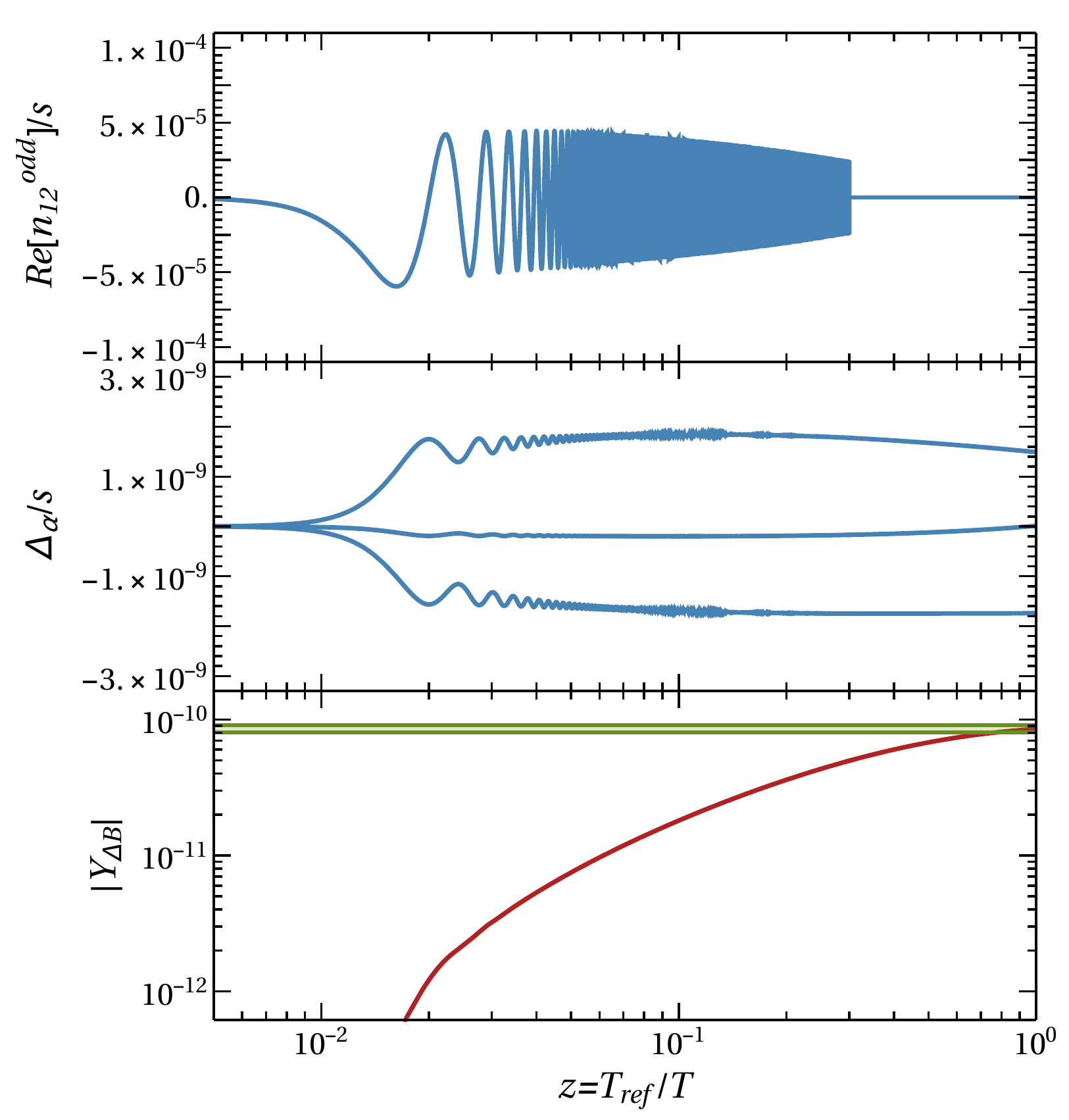}
	\includegraphics[width=0.495\textwidth]{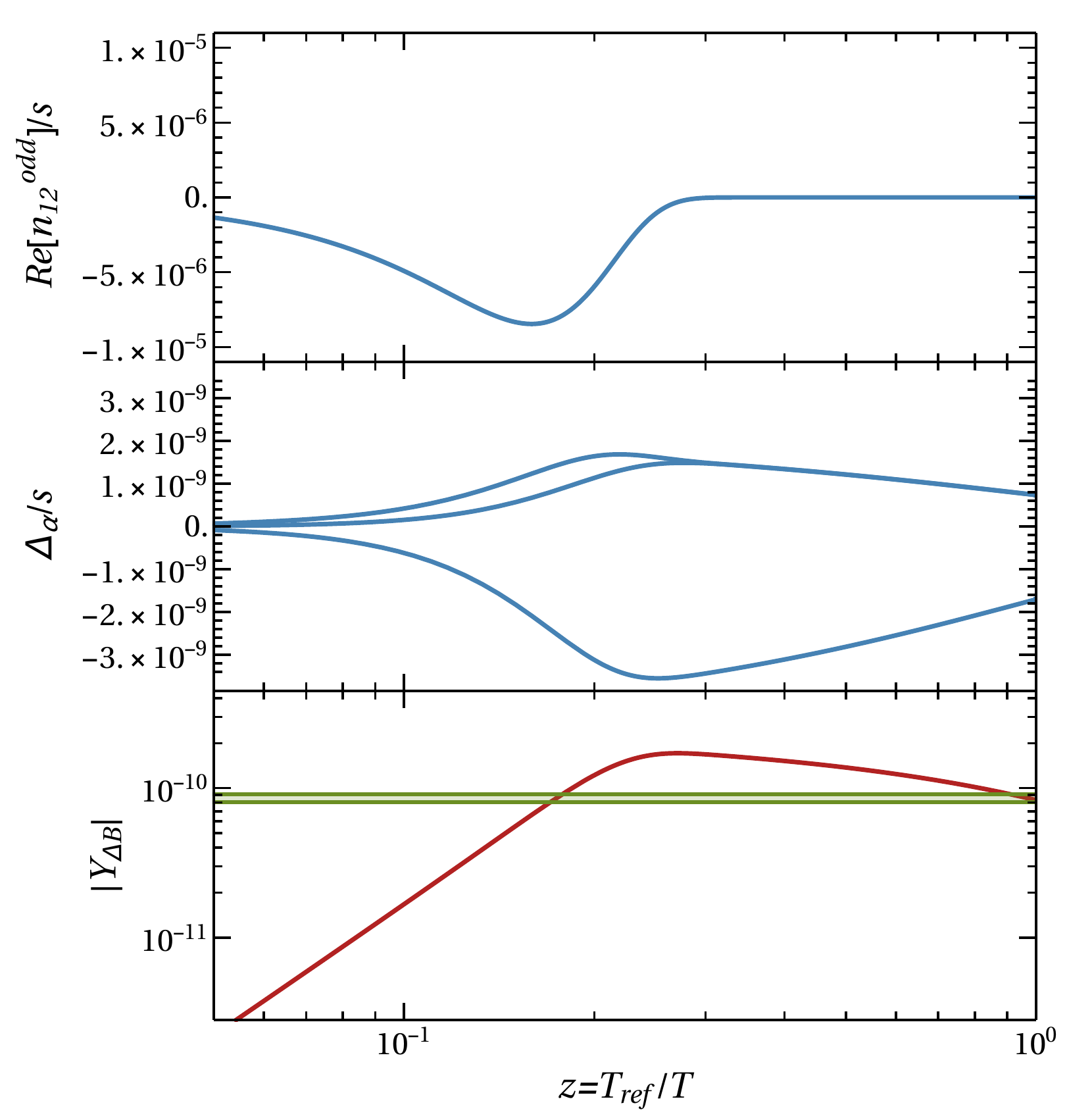}
\caption{\label{fig:oscillatory:overdamped}Evolution of the helicity-odd density of RH neutrino correlations $\delta n_{12}^{\rm odd}$,
the SM conserved asymmetries $\Delta_a$, $a=e,\mu,\tau$ and the baryon-to-entropy ratio
$Y_{\Delta B}$ over $z=T_{\rm ref}/T$ for $T_{\rm ref}=130\,{\rm GeV}$. For the panel on the left,
an exemplary parameter point in the oscillatory regime is chosen, and on the right a point
representing the overdamped regime.}
\end{figure}

\subsubsection{Oscillatory regime}

The dynamics of the oscillatory regime is exemplified in the left panel of \fref{fig:oscillatory:overdamped}. The approach for analytic approximations
is first to consider the evolution of the RH neutrinos through the first
few oscillations while neglecting the backreaction from the doublet lepton asymmetries.
These early asymmetries can then rather accurately be computed from a source term
that depends on the phase oscillations in the correlations of the different mass eigenstates of the RH neutrinos. In a second step, at later times toward electroweak symmetry breaking, we consider the redistribution of the leptonic charges that
then induces in interplay with sphaleron processes a net baryon asymmetry.\footnote{If one of the active flavors only couples very weakly to the RH neutrinos, washout of the
pertaining asymmetry can be avoided to a large extent. This leads to a variant of ARS that
also works with RH neutrinos above the electroweak scale~\cite{Garbrecht:2014bfa}.}

\paragraph{Oscillations and creation of early asymmetries}
When the oscillations at temperatures way above the electroweak scale are fast compared
to the Hubble rate (i.e. the mass-degeneracy is moderate at most), thermal corrections to the RH neutrino mass are negligible. Moreover, if the RH neutrinos are far from equilibrium, we
can neglect the backreaction effects from the doublet leptons such that we can simplify Eq.~\eqref{diff:sterile} to
\begin{align}
\label{osc:weakwashoutav}
\frac{\rm d}{{\rm d} z}\delta n_{h} +\frac{i}{2}z^2[H_N^{\rm vac},\delta n_{h}]=-\frac{1}{2}\{\Gamma_N,\delta n_{h}\}\,.
\end{align}
Then, assuming vanishing initial distributions of the RH neutrinos, we can approximate the
diagonal number densities at early times by
\begin{align}
\delta n_{h\,ii}&=-n^{\rm eq}+\mathcal{O}(|\lambda^T \lambda^*|)\,.
\end{align}
When we substitute this on the right-hand side of Eq.~\eqref{osc:weakwashoutav},
we obtain the following equations for the
off-diagonal number densities ($i\not=j$):
\begin{subequations}
\begin{align}
\label{off_diagonal_de}
\frac{\rm d}{{\rm d}z}n^{\rm odd}_{ij}+{i} A_{ij} z^2 n^{\rm odd}_{ij}&=-{i} {\rm Im}[\lambda^T \lambda^*]_{ij}G\,,\\
\frac{\rm d}{{\rm d}z}n^{\rm even}_{ij}+{i} A_{ij} z^2 n^{\rm even}_{ij}&={\rm Re}[\lambda^T \lambda^*]_{ij}G\,,
\end{align}
\end{subequations}
where
\begin{align}
A_{ij}=\frac{a_{\rm R}}{T_{\rm ref}^3}\frac{\pi^2}{36 \zeta(3)}(-\Delta M_{N ij}^2)\,, \quad \quad
G=\gamma_{\rm av}\frac{a_{\rm R}}{T_{\rm ref}}
n^{\rm eq}\,.
\end{align}
For vanishing initial abundances of the RH neutrinos,
the solutions to these equations are
\begin{subequations}
\label{sol:f0hav}
\begin{align}
n^{\rm odd}_{ij}&=-{i}{\rm Im}[\lambda^T \lambda^*]_{ij}G\mathcal{F}_{ij}\,,\quad \quad 
n^{\rm even}_{ij}={\rm Re}[\lambda^T \lambda^*]_{ij}G\mathcal{F}_{ij}\,,\\
\label{F_tilde}
\mathcal{F}_{ij}&=\left[\frac{\Gamma\left(\frac{1}{3}\right)}{3^{\frac 23}(-{i} A_{ij})^{\frac 13}}-\frac{z}{3}E_{2/3}\left(-\frac{i}{3} A_{ij} z^3\right)\right]\exp \left(-\frac{{i}}{3} A_{ij}z^3\right)\,,
\end{align}
\end{subequations}
where, when using the rates derived in \sref{sec:non-equlibrium},
\begin{align}
E_n(x)=\int\limits_1^\infty \!{\rm d} t\,\frac{{\rm e}^{-xt}}{t^n}\,.
\end{align}
It is further possible to show that up to order $|\lambda^\dagger\lambda|$,
no sterile charges are produced in the diagonal components of $\delta n^{\rm odd}$ (for $n_R=2$ RH neutrino flavors, this holds to all orders),
such that we can neglect the approximately conserved fermion number in the
RH neutrino sector to this end.

The charges $\Delta_\alpha$ reach their maximal values $\Delta_\alpha^{\rm sat}$
after only a few oscillations beyond which the oscillatory,
CP-violating source averages out, cf. \fref{fig:oscillatory:overdamped}.
We can therefore approximate
\begin{align}
\Delta_\alpha(z)= -\int_0^z  \!\frac{{\rm d} z'}{T_{\rm ref}}S_{\alpha}\approx -\int_0^\infty \!\frac{{\rm d} z'}{T_{\rm ref}}S_{\alpha}\equiv \Delta_{\alpha}^{\rm sat}\,,
\end{align}
which holds for $z$ moderately larger than $z_{\rm osc}$.
Substituting the approximate solution~\eqref{sol:f0hav} into
the source term~\eqref{Source} and evaluating the integral above, we obtain
\begin{align}
\label{sol:act_av}
\nonumber
\frac{\Delta_\alpha^{\rm sat}}{s}&=\frac{i}{g_\star^{\frac 53}}
\frac{3^{\frac{13}{3}} 5^{\frac53}\Gamma(\frac16)\zeta(3)^{\frac53}}{2^{\frac{8}{3}}\pi^{\frac{41}{6}}}
\sum\limits_{\overset{i,j,c}{i\not=j}}
\frac{
\lambda_{\alpha i} \lambda^\dagger_{i\gamma} \lambda_{\gamma j} \lambda^\dagger_{j\alpha}
}{{\rm sign}(\Delta M_{Nji}^2)}\left(\frac{M_{\rm Pl}^2}{|\Delta M_{Nij}^2|}\right)^{\frac23}\frac{\gamma_{\rm av}^2}{g_w}\\
&\approx
-\sum\limits_{\overset{i,j,c}{i\not=j}}
\frac{\Im[\lambda_{\alpha i} \lambda^\dagger_{i\gamma} \lambda_{\gamma j} \lambda^\dagger_{j \alpha}]}{{\rm sign}(\Delta M_{Nji}^2)}
\left(\frac{M_{\rm Pl}^2}{|\Delta M_{Nij}^2|}\right)^{\frac23}
%\times 3.39392
\times 3.4
\times 10^{-4}\frac{\gamma_{\rm av}^2}{g_w}\,.
\end{align}
This asymmetry is yet purely flavored such that it is crucial now to incorporate the
washout effects that allow for a net baryon-to-entropy ratio $Y_{\Delta B}$ to
emerge at late times, cf. \fref{fig:oscillatory:overdamped}.

\paragraph{Washout and redistribution of leptonic charges}
By definition,
washout begins in earnest when $z\sim z_{\rm eq}$, where the off diagonal correlations of the RH neutrinos have already decayed or they are oscillating so rapidly that their effect
averages out. We can therefore neglect the source term for the asymmetry as
well as the off-diagonal RH neutrino correlations, such that
the fluid equations can 
at this stage be approximated by
\begin{subequations}
\label{eq:washout}
\begin{align}
\label{eq:washout_1}
\frac{{\rm d} \Delta_\alpha}{{\rm d} z}&=
\frac{\gamma_{\rm av}}{g_w} \frac{a_{\rm R}}{T_{\rm ref}}\sum_{i} \lambda^\dagger_{i\alpha}\lambda_{\alpha i}
\,\left( \sum_{b}(-C_{\alpha\beta} + B_\beta/2)\Delta_\beta-q_{Ni}\right)\,,
\\
\label{eq:washout_2}
\frac{{\rm d} q_{Ni}}{{\rm d} z}&=-\frac{a_{\rm R}}{T_{\rm ref}}\gamma_{\rm av}
\sum\limits_a \lambda^\dagger_{i\alpha}\lambda_{\alpha i}
\left(q_{Ni}-\sum_b (-C_{\alpha\beta} + B_\beta/2)\Delta_\beta\right)\,,
\end{align}
\end{subequations}
where $\Delta_{\alpha}^{\rm sat}$  and $q_N=0$ are the initial conditions for $z\to 0$.
Equation~(\ref{eq:washout_1}) is the same as Eq.~(\ref{evolution:active}) without the source term, and Eq.~(\ref{eq:washout_2}) is derived from  Eq.~(\ref{diff:sterile}) where we
only keep the decay term $\Gamma_N$ as well as the backreaction term $\tilde{\Gamma}_N$
and take the helicity-odd combination.
For $n_R$ RH neutrinos we recast  Eqs.~(\ref{eq:washout}) to a
system of
linear first-order differential equations for a $(3+n_R)$-dimensional vector $V_{\Delta N}=(\Delta^t, q_N^t)^t$,
\begin{align}
\frac{{\rm d}}{{\rm d} z}
V_{\Delta N}=\frac{a_{\rm R}}{T_{\rm ref}}\gamma_{\rm av}
K V_{\Delta N}
\,, \quad \quad K=\left(
\begin{array}{ccc}
K^{\Delta \Delta} & K^{\Delta N} \\
K^{N \Delta} & K^{N N} 
\end{array}
\right)\,.
\end{align}
The matrices $K^{\Delta \Delta}, K^{\Delta N}, K^{N \Delta}$ and  $K^{N N}$ in their
components read
\begin{align}
\nonumber
K^{\Delta \Delta}_{\alpha\beta}&=\frac{1}{g_w}\sum_{k=1}^{n_s} \lambda_{\alpha k} \lambda^\dagger_{ka}(-C_{\alpha\beta}+\frac{1}{2}E_{\alpha\beta})\,, \quad  \quad
K^{\Delta N}_{\alpha j}=-\frac{1}{g_w} \lambda_{\alpha j} \lambda^\dagger_{j\alpha}\,,\\
K^{N \Delta}_{i\beta}&=\sum_{\delta=1}^3\lambda^\dagger_{i\delta}\lambda_{\delta i} (-C_{\delta\beta}+\frac{1}{2}C_\beta)\,,\quad \quad\quad\quad\,
K^{N N}_{ij}=-\sum_{\delta=1}^3 \lambda_{i\delta}^\dagger\lambda_{\delta i} \delta_{ij}\,,
\end{align}
with $i,j=1,2,\dots,n_R$ RH neutrinos and $\alpha,\beta=1,2,3$ flavors of active leptons.
The matrix $E_{ab}$ is $3\times 3$ with $1$ in each entry, and $C$ and $B$ are defined in Eqs.~(\ref{linrel:spectators},\ref{linrel:spectators:2}) account for the spectator processes.
When diagonalizing
\begin{align}
\label{wo:diag}
K^{\rm diag}={\rm T}^{-1}K {\rm  T}\,,
\end{align}
we can write down the solution
\begin{align}
\left(
\begin{array}{ccc}
\Delta(z) \\
q_N(z)
\end{array}
\right)={\rm T}\,\exp\left(\frac{a_{\rm R}}{T_{\rm ref}}\gamma_{\rm av}  K^{\rm diag} \,z\right){\rm T}^{-1}
\left(
\begin{array}{ccc}
\Delta^{\rm in} \\
q_N^{\rm in}
\end{array}
\right)\,,
\end{align}
where $\Delta^{\rm in}=\Delta^{\rm sat}$ and $q_{N}^{\rm in}=0$ are
the asymmetries that are generated during the oscillation process at early times
$z\sim z_{\rm osc}$ and that we impose as the boundary conditions to Eqs.~\eqref{eq:washout}.
Eventually, the density of baryon charge $B$ freezes in when weak sphalerons are quenched, which occurs for $z=1$ in our parametrization. The baryon number density then takes the value
\begin{align}\label{BapproxOscillating}
B=\frac{28}{79}
[\Delta_1(z)+\Delta_2(z)+\Delta_3(z)]_{z=1}\,,
\end{align}
and $Y_{\Delta B}=B/s$.
The results discussed here agree up to the numerical prefactor with
what is derived in \sref{sec:expansion:mixings} with the result of Eq.~\eqref{asymmetry:Valencia:perturbative} for the weak washout regime.
Since we do not rely here on the expansion in small mixing
angles, the result shown here is somewhat implicit because of
the diagonalization~\eqref{wo:diag}, which however is performed easily
at little numerical cost.

\subsubsection{Overdamped regime}
\label{sec:overdamped}

The overdamped regime where $z_{\rm eq}\ll z_{\rm osc}$ requires
a strong level of mass-degeneracy among the RH neutrinos or large RH neutrino
Yukawa couplings compared to the smallest viable values in the seesaw parameter
space. In order to avoid strong parametric tuning, one may appeal
to lepton number as an approximately conserved classical symmetry~\cite{Mohapatra:1986bd,Mohapatra:1986aw,GonzalezGarcia:1988rw,Branco:1988ex}, where two of the
RH neutrino states can be arranged to form a pseudo-Dirac fermion. For $n_R=2$ RH neutrinos,
the case for which we here discuss the analytic approximations, the overdamped regime
implies approximate lepton number conservation.

A salient feature of approximate lepton number conservation is that
one of the \emph{interaction} eigenstates of the RH neutrinos almost decouples.
In the interaction basis, where $\lambda^\dagger\lambda$ is diagonal and
\begin{subequations}
\begin{align}
    \Gamma_N &=
	\gamma_{\rm av} \frac{a_{\rm R}}{T_{\rm ref}}
	\begin{pmatrix}
	(\lambda^\dagger\lambda)_{11} & 0 %& 0
	\\0 & (\lambda^\dagger\lambda)_{22}
    \end{pmatrix}
 \,,\\
	H_N^{\rm th} &=
	(\mathfrak{h}_{\rm th}+\mathfrak{h}_{\rm EV}(z)) \frac{a_{\rm R}}{T_{\rm ref}}
	\begin{pmatrix}
	(\lambda^\dagger\lambda)_{11} & 0 
	\\ 0 & (\lambda^\dagger\lambda)_{22}
    \end{pmatrix}\,,
\end{align}
\end{subequations}
we can therefore neglect the smaller eigenvalue, i.e. the second one
for definiteness, and drop all terms of 
$\mathcal{O}\left((\lambda^\dagger\lambda)_{22}\right)$.
In the interaction basis,
the Hamiltonian due to the
vacuum mass matrix $H_N^{\rm vac}$ is non-diagonal in general, i.e.
it takes the form
\begin{align}
    H_N^{\rm vac}=
	\frac{\pi^2}{18 \zeta(3)}\frac{a_{\rm R}}{T_{\rm ref}^3}
    \begin{pmatrix}
		M^2_{N 11} & M^2_{N 12}
		\\ {M_{N 12}}^{\!\!\!\!\!\!*2} & M_{N 22}
    \end{pmatrix}\,.
\end{align}
In this setup, the condition for being in the overdamped regime reads
\begin{align}
	\frac{z_{\rm eq}}{z_{\rm osc}} =
	\frac{\sqrt[3]{|M^2_{1}-M^2_{2}|/a_{\rm R}^2}}
	{\gamma_{\rm av}(\lambda^\dagger\lambda)_{11}}
	\ll 1\,.
\end{align}

The fluid equations for the RH neutrinos then immediately decompose into
a set involving direct damping through
$(\lambda^\dagger\lambda)_{11}$,
\begin{subequations}
\label{osc2x2:equilibrated}
\begin{align}
\frac{{\rm d} \delta n_{11}}{{\rm d} z} &= - (\Gamma_N)_{11} \delta n_{11}
	-\frac{{i}}{2} z^2
	\left[ (H_N^{\rm vac})_{12} \delta n_{21} - (H_N^{\rm vac})^*_{12} \delta n_{12}\right]\,,\\
	\frac{{\rm d} \delta n_{12}}{{\rm d} z} &= - \frac{(\Gamma_N)_{11}}{2} \delta n_{12}
	- {i} \frac{(H_N^{\rm th})_{11}}{2} \delta n_{12}
    - \frac{{i}}{2} z^2 \sum_k \left[ (H_N^{\rm vac})_{1k} \delta n_{k2} -
	\delta n_{1k} (H_N^{\rm vac})_{k 2} \right]\,,
\end{align}
\end{subequations}
and one equation where damping occurs indirectly, through mixing with other sterile flavors,
\begin{align}
\label{osc2x2:noneq}
 \frac{{\rm d} \delta n_{22}}{{\rm d} z} &= - \frac{{i}}{2} z^2 \left[ (H_N^{\rm vac})^*_{12} \delta n_{12} -
	 (H_N^{\rm vac})_{12} \delta n_{21}\right]\,.
\end{align}
Here, we have suppressed the helicity index $h$, which we only show if necessary
in the following.

The key approximation that we apply now is one that is familiar from
resonant leptogenesis in the strong washout regime~\cite{Iso:2014afa,Garbrecht:2014aga,leptogenesis:A03}.
Due to the hierarchy in the RH neutrino Yukawa couplings, the more strongly coupled
RH neutrino $N_1$ as well as its flavor correlations with $N_2$ almost instantaneously
reach the value that is implied by the non-vanishing $\delta n_{22}$.
Thus, we neglect the derivative terms in Eqs.~\eqref{osc2x2:equilibrated} and solve
these algebraically for $\delta n_{ij}$ with $i\not=j$. These solutions are then substituted
back into Eq.~\eqref{osc2x2:noneq} that then determines the time-dependent dynamics. 
This procedure is
appropriate in particular if the rates implied on the right-hand side of Eqs.~\eqref{osc2x2:equilibrated} are larger than the Hubble rate. Somewhat tedious
but straightforward algebra thus leads to the CP-odd correlation
\begin{align}
	\delta n_{+\,12} - \delta n_{-\,12}^* = - \frac{2 z^2 {i} (H_N^{\rm vac})_{12} (\Gamma_N)_{11}}
	{(\tilde{H}_N^{\rm vac})^2 (z^2+\tilde{z}_{\rm c}^2)(z^2+\tilde{z}_{\rm c}^{*2})} \delta n_{22}(z)\,,
\label{osc2x2:n12cpo}
\end{align}
that enters the source term~\eqref{Source} as
\begin{align}
\label{eq:source_sw}
\nonumber
	S_{\alpha} &= a_{\rm R} \frac{\gamma_{\rm av}}{g_w}
	\sum\limits_{\overset{i,j}{i\not=j}} \lambda_{\alpha i} \lambda^\dagger_{j\alpha} \left( \delta n_{+\,ij} - \delta n^*_{-\,ij} \right)\\ 
	&=4 
	\frac{\gamma_{\rm av}^2 a_{\rm R}^2}{g_w T_{\rm ref}}
	\frac{\sum_\beta |\lambda_{\beta 1}|^2}{(\tilde{H}_N^{\rm vac})^2}\frac{z^2}{|z^2+\tilde{z}_{\rm c}^2|^2}
	{\rm Im}\left[\lambda_{\alpha 1} (H_N^{\rm vac})_{12} \lambda^\dagger_{2\alpha} \right]\delta n_{22}(z)
	\,.
\end{align}
Here, we have introduced
\begin{align*}
	(\tilde{H}_N^{\rm vac})^2 \equiv|(H_N^{\rm vac})_{12}|^2 +
	\left[ (H_N^{\rm vac})_{11}- (H_N^{\rm vac})_{22} \right]^2
\end{align*}
and
\begin{align}
	\tilde{z}_{\rm c} = \sqrt{\frac{(H_N^{\rm th})_{11}}{\tilde{H}_N^{\rm vac}}
	\left[ \frac{(H_N^{\rm vac})_{11}-(H_N^{\rm vac})_{22}}{ \tilde{H}_N^{\rm vac}}+
		{i} \sqrt{\frac{|(H_N^{\rm vac})_{12}|^2}{(\tilde{H}_N^{\rm vac})^2}+
\frac{\gamma_{\rm av}^2}{\mathfrak{h}_{\rm th}^2}}\right]}\,.
\end{align}
Moreover, the approximate solution for the number density of the more weakly coupled RH neutrino is
\begin{align}
\label{deltan22}
	\delta n_{22}= \delta n_{22}(0) \exp\left\{-(\Gamma_N)_{11} \frac{|(H_N^{\rm vac})_{12}|^2}
	{(\tilde{H}_N^{\rm vac})^2}
    \left[z-\frac{\Im \left(\tilde{z}_{\rm c}^3 \arctan \frac{z}{\tilde{z}_{\rm c}}\right)}{\Im\tilde{z}_{\rm c}^2}\right]\right\}\,.
\end{align}

While in the oscillatory regime, the early oscillations of the off-diagonal
correlations of the RH neutrinos lead to a separation of the production from the washout
of SM charges, in the overdamped regime, both aspects
have to be considered simultaneously because these are driven by the damping mediated
through the RH neutrino Yukawa couplings, cf.~\fref{fig:oscillatory:overdamped}.

In the present setup, we assume that the processes mediated by the larger Yukawa couplings
$\lambda_{\alpha 1}$ are fast compared to the Hubble rate, i.e. $N_1$ can be treated
as a fully equilibrated spectator. This modifies the evolution of the SM conserved
charges [cf. Eq.~\eqref{evolution:active}] to
\begin{align}
\label{Da:sw}
	\frac{{\rm d} \Delta_\alpha}{{\rm d} z} &= \sum_\beta \tilde{W}_{\alpha\beta} \Delta_\beta 
	- \frac{S_{\alpha}(z)}{T_{\rm ref}} - \frac{2}{g_w}\frac{|\lambda_{\alpha1}|^2}{(\lambda^\dagger\lambda)_{11}} \frac{{\rm d} \delta n_{22}^{\rm odd}}{{\rm d} z}\,,
\end{align}
where
\begin{align}
	\tilde{W}_{\alpha\beta}&=-\frac{a_{\rm R}}{T_{\rm ref}}\frac{\gamma_{\rm av}}{g_w}
	|\lambda_{\alpha 1}|^2
	\sum_\gamma \left( \delta_{\alpha\gamma} - \frac{|\lambda_{\gamma 1}|^2}{(\lambda^\dagger\lambda)_{11}} \right)C_{\gamma\beta}\,.
\end{align}
Given the solution~\eqref{deltan22}, Eq.~\eqref{Da:sw} can simply be integrated. Making use
of the approximate fermion number conservation in the limit of small Majorana masses
and the fact that $N_1$ acts as a spectator, we can further derive for the baryon number density
\begin{align}\label{BapproxOverdamped}
	B(z)\approx \frac{28}{79}
	\left[
		\sum_{\alpha\beta} \Delta_{\alpha}(z) (-C_{\alpha\beta}+B_\beta/2)
		\frac{|\lambda_{\alpha 1}|^2}{g_w [\lambda^\dagger\lambda]_{11}}
		+\frac{2}{g_w}\delta n_{22}^{\rm odd}(z)
	\right]\,.
\end{align}

We emphasize that because the overdamped regime allows for large active-sterile mixing angles while
being in agreement with the observed oscillation and mass-scales of the light
active neutrinos, it is of particular relevance in view of direct search bounds
for RH neutrinos that are available currently as well as in the intermediate future.
In particular, for the parametric configurations corresponding to the
upper bound on the active-sterile mixing that is
shown in \fref{fig:U2maxNHIH}, ARS leptogenesis takes place in the overdamped regime.

\section{Numerical results}
\label{section:numerical}

The parameter space on which these models can explain both the measured neutrino masses and mixings, as well as the matter-antimatter asymmetry has been studied in several 
works~\cite{Hernandez:2016kel,Drewes:2016gmt,Drewes:2016jae,Canetti:2012zc,Canetti:2012kh,Abada:2015rta,Hernandez:2015wna,Shuve:2014zua,Drewes:2012ma}. Particularly interesting is  to know the range of masses of the heavy neutrinos and their mixings to the charged leptons, since both are measurable quantities, as well as 
the possible correlation of the matter-antimatter asymmetry with the CP-violating phases in the neutrino mixing matrix, or the amplitude of neutrinoless double-beta decay. 
We describe the results of two recent scans of parameter space in the minimal model. 

\subsection{Bayesian analysis in the  minimal model}

A state-of-the-art Bayesian analysis has been performed\cite{Hernandez:2016kel} in the minimal model, $n_R=2$. 
Defining the log-likelihood
\begin{eqnarray}
\log {\mathcal L} = -{1\over 2} \left({Y_{\Delta B}(t_{\rm EW})-Y^{\rm exp}_{\Delta B}\over \sigma_{Y_{\Delta B}}}\right)^2,
\end{eqnarray}
a nested sampling algorithm implemented in the public package MultiNest
\cite{Feroz:2007kg,Feroz:2008xx} and the Markov Chain sample analysis tool
GetDist are used to get posterior probabilities. 
At each point of the parameter space sampled the kinetic Eqs.~\eqref{eq:rhonrhonbarav} are solved numerically using the software SQuIDS \cite{Delgado:2014kpa}, designed to solve the evolution of a generic density matrix in the interaction picture. For further details on the numerical treatment of these equations we refer to Ref \cite{Hernandez:2016kel}. 

The scan is performed using the Casas-Ibarra parameters of Eq.~\eqref{eq:yci}.  The light neutrino masses and mixings are fixed to the present best fit points 
in the global analysis of neutrino oscillation data of Ref.~\cite{Gonzalez-Garcia:2014bfa}, for each of the neutrino orderings (normal, NH, and inverted, IH), and the remaining free parameters are: the complex angle of the $R$ matrix, the CP phases of the PMNS matrix and the heavy Majorana masses.  Flat priors are assumed for all the angular parameters, for $\log_{10}\left({ M_{1}\over{\rm GeV}}\right)$, within the range $M_{1} \in [0.1{\rm GeV} , 10^2 {\rm GeV}]$, and  for the second mass two priors are considered: 1) a flat prior also in $\log_{10} \left({M_{2}\over GeV}\right)$ in the same range or 2) a flat prior in $\log_{10} \left({|M_{2}-M_{1}|\over {\rm GeV}}\right)$ in the range $M_{2}-M_{1} \in [10^{-8}{\rm GeV} , 10^2{\rm GeV}]$. 

Figures~\ref{fig:trianglen2ih} and \ref{fig:trianglen2nh} show, for IH and NH, the posterior probabilities of the spectrum of the two relevant states, $M_{1}, M_{2}$, the active-sterile mixings of the first heavy state $|U_{\alpha 4}|^2$ (those of the second state are almost identical), the neutrinoless double beta decay effective mass  $|m_{\beta\beta}|$ and the baryon asymmetry $Y_{\Delta B}$. The two different colors (light blue and red) correspond to the two prior options for $M_{2}$.  The significant differences between the two posteriors show the effect of allowing or not for fine-tuning in the degeneracy of the two heavy states. Even though the contours are typically larger if more fine-tuning is allowed,  interesting solutions with a mild degeneracy are found, which tend to correlate with smaller $M_{1}, M_{2}$,  larger values of the active-sterile mixing parameters and a sizable non-standard contribution to neutrinoless double beta decay, which obviously imply much better chances of testability. The chances of testability and the correlation with CP violation in mixing will be discussed in more detail in the accompanying review article~\cite{leptogenesis:A05}.
 \begin{figure}[!ht]
 \begin{center}

\includegraphics[scale=0.3]{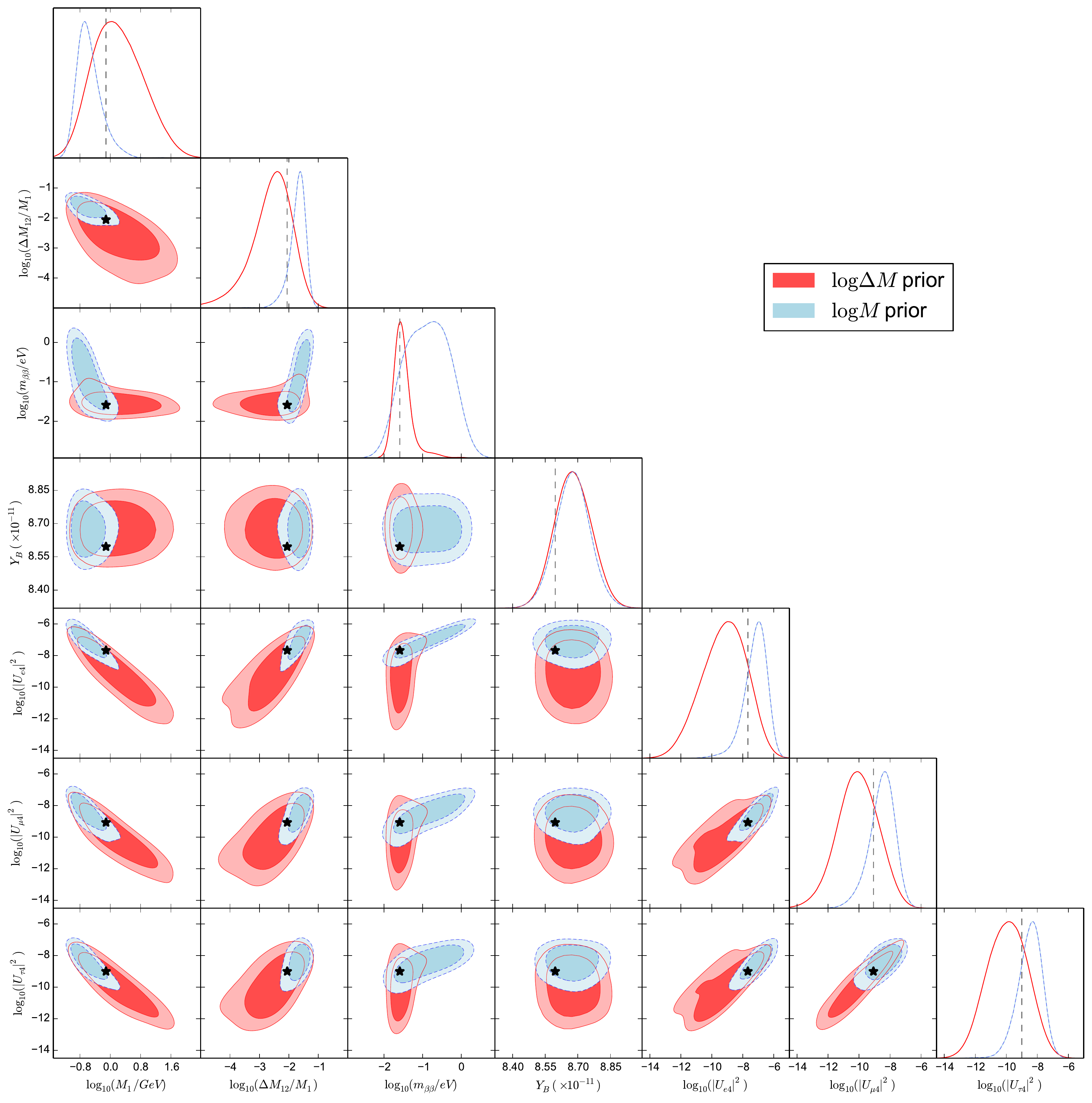} 
\caption{\label{fig:trianglen2ih} Triangle plot with 1D posterior probabilities and  2D $68\%$ and $90\%$ probability contours in the $n_R=2$ scenario for IH. The parameters shown are the observables $M_{1}$, $\Delta M_{N 12}/M_{1}=(M_{2}-M_{1})/M_{1}$, $m_{\beta\beta}$, $Y_{\Delta B}$,  and the three mixings with the first of the heavy states $|U_{\alpha 4}|^2$ for $\alpha=e,\mu, \tau$. The blue and red contours correspond respectively to the assumption of a flat prior in $\log_{10} M_{1}$ and $\log_{10} M_{2}$ and to a flat prior in $\log_{10} 
M_{1}$ and $\log_{10}(\Delta M_{N 12})$.  Plot taken from Ref.~\cite{Hernandez:2016kel}. 
 }
\end{center}
\end{figure}

 \begin{figure}[!ht]
 \begin{center}

\includegraphics[scale=0.3]{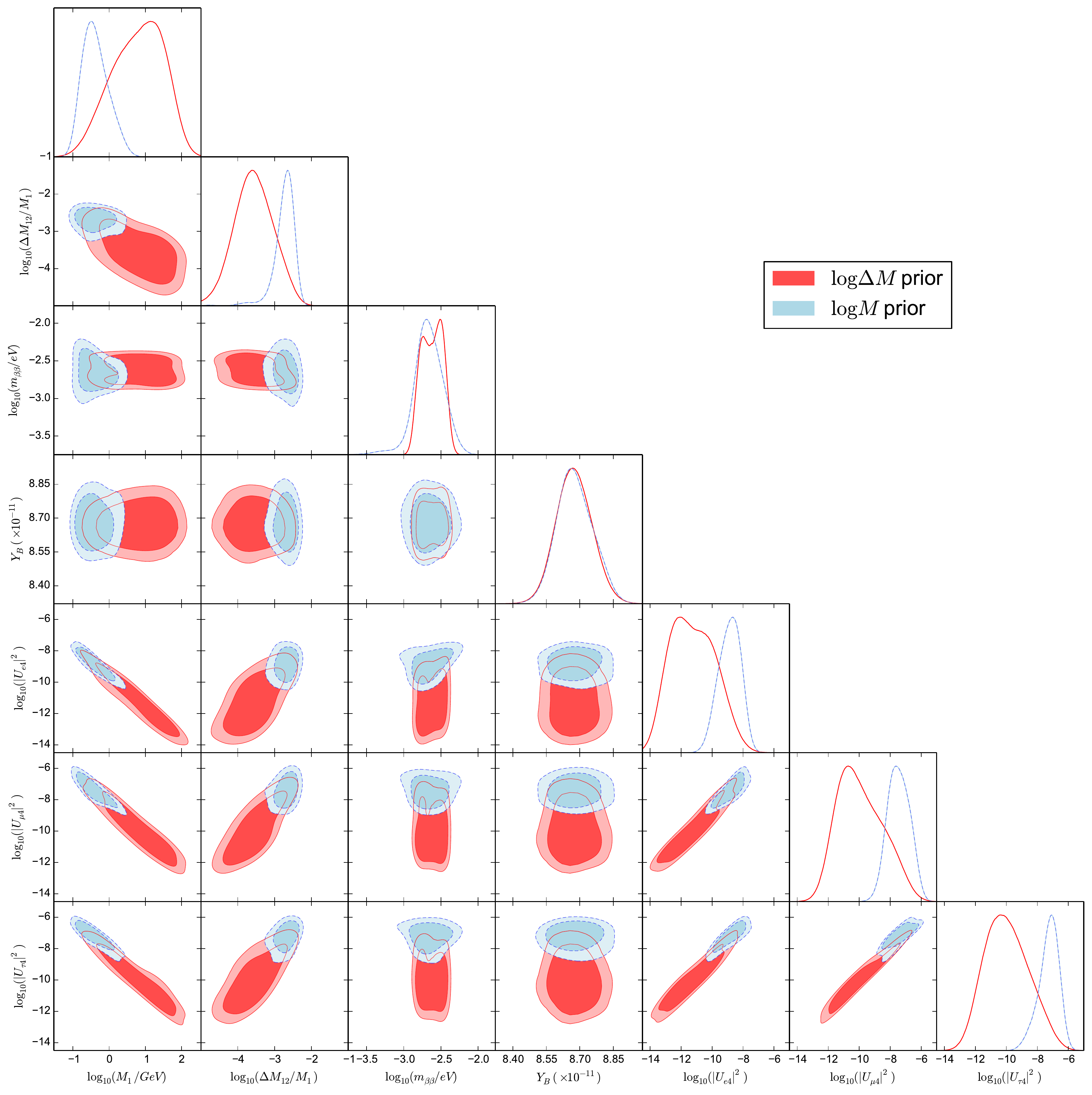} 
\caption{\label{fig:trianglen2nh} Triangle plot with 1D posterior probabilities and  2D $68\%$ and $90\%$ probability contours in the $n_R=2$ scenario for NH. The parameters shown are the observables $M_{1}$, $\Delta M_{N 12}/M_{1}=(M_{2}-M_{1})/M_{1}$, $m_{\beta\beta}$, $Y_{\Delta B}$,  and the three mixings with the first of the heavy states $|U_{\alpha 4}|^2$ for $\alpha=e,\mu, \tau$. The blue and red contours correspond respectively to the assumption of a flat prior in $\log_{10} M_{1}$ and $\log_{10} M_{2}$ and to a flat prior in $\log_{10} 
M_{1}$ and $\log_{10}(\Delta M_{N 12})$. Plot taken from Ref.~\cite{Hernandez:2016kel}. 
 }
\end{center}
\end{figure}

\subsection{Large-mixing solutions in the minimal model}
\label{sec:largemix}

Approximate lepton-number conservation~\cite{Mohapatra:1986bd,Mohapatra:1986aw,GonzalezGarcia:1988rw,Branco:1988ex} for $n_R=2$ RH neutrinos (as well as for $n_R>2$ provided the
additional RH states decouple) implies a strong mass degeneracy among
the pseudo-Dirac RH neutrino pair and allows for larger Yukawa couplings and consequently for
large active-sterile mixing.
As it is explained in \sref{sec:overdamped}, ARS leptogenesis then tends to
take place in the overdamped regime, in particular when the active-sterile mixing
is close to the maximal presently allowed level. The dynamics of this
regime is illustrated in the right panel of~\fref{fig:oscillatory:overdamped}.
 
\begin{figure}[!ht]
\begin{center}
\includegraphics[width=0.8\textwidth]{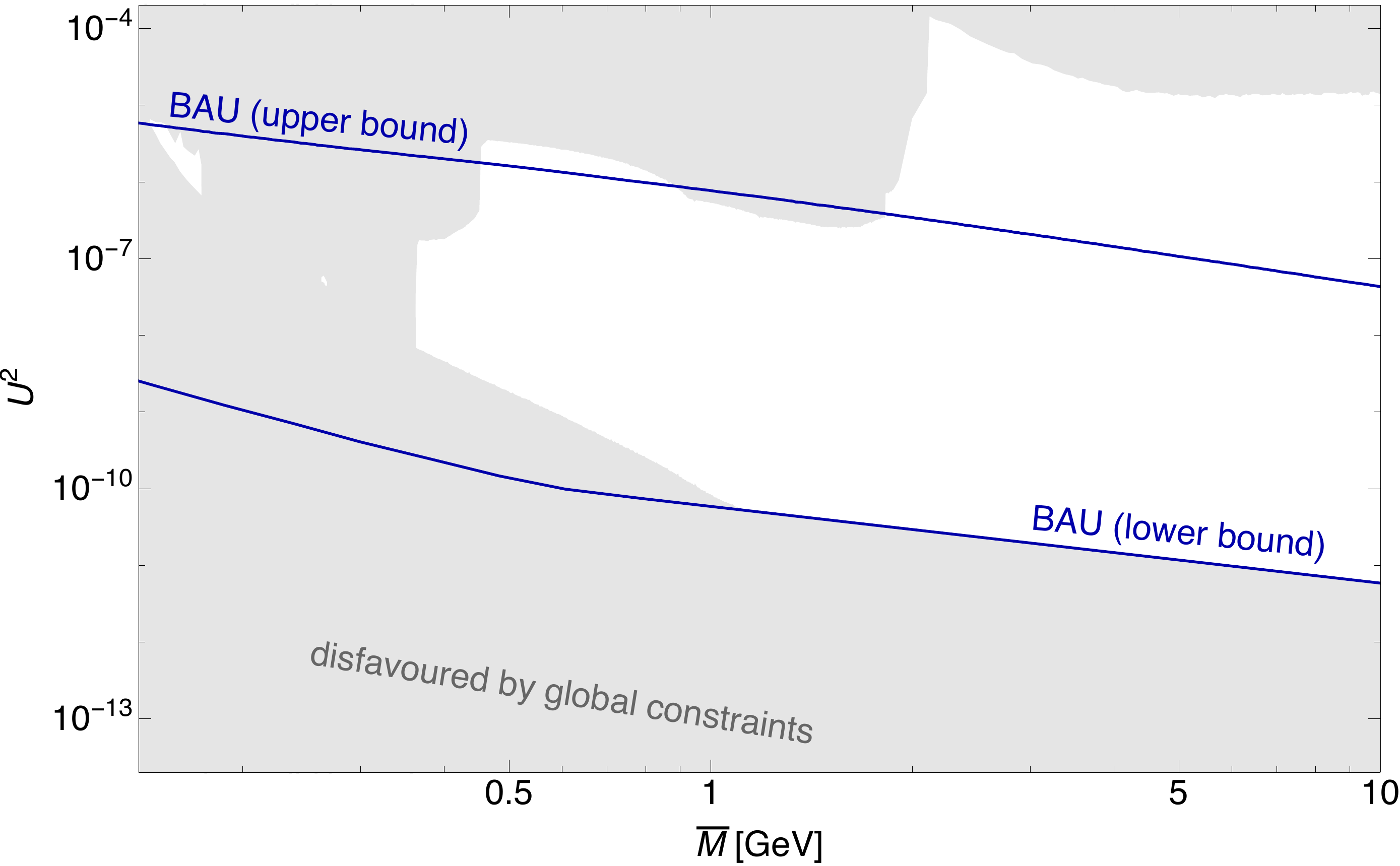}
\includegraphics[width=0.8\textwidth]{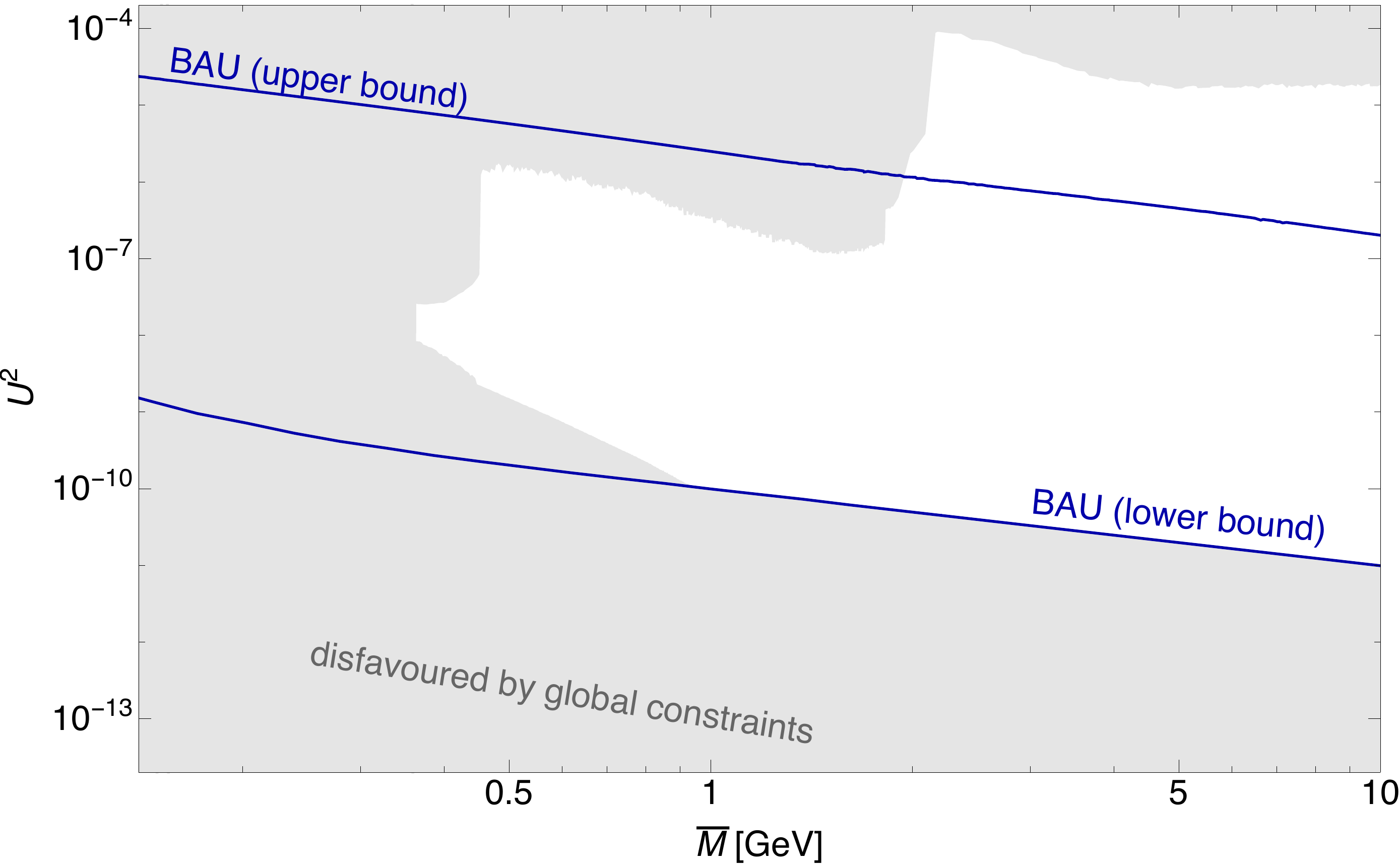}
\caption{\label{fig:U2maxNHIH}The viable range for ARS leptogenesis is shown between the blue lines
for normal hierarchy in the upper panel and for inverted hierarchy in the lower panel.
Superimposed are
exclusions from past experiments and big bang nucleosynthesis (BBN) along with the theoretical lower
bound on the mixing from the seesaw mechanism. We assume $n_R=2$ RH neutrinos, $\bar M$ is their average mass, and
$U^2=\sum_{i,\alpha} |U_{\alpha i}|^2$ the total active-sterile mixing.}
\end{center}
\end{figure}

In order to determine the upper bound on the active-sterile mixing,
one can make use of the insights from \sref{sec:overdamped} into
the analytic dependencies of the baryon asymmetry on the input parameters as
well as some general scaling arguments~\cite{Drewes:2016gmt}.
First, the flavored asymmetries are most efficiently turned into a net baryon asymmetry
provided one of the active flavors (i.e. the one that mixes most
strongly with the lightest of the light neutrino mass eigenstates) couples
as weakly as possible to the RH states. This readily fixes the remaining CP phases
in the PMNS matrix that are yet to be determined experimentally. (In turn, in the case of a discovery of RH neutrinos with large active-sterile mixing, this
leads to a particular opportunity of testing ARS leptogenesis with $n_R=2$
sterile flavors, see Ref.~\cite{Drewes:2016jae} and the pertaining discussion in the
accompanying paper~\cite{leptogenesis:A05}). Then, maximizing the
CP-violating source also fixes the real part of the complex
Casas-Ibarra angle. Finally, it can be shown that a rescaling
$M_N\to \xi M_N$ and $\Delta M_N^2 \to \eta\Delta M_N^2$ results in a change
of the final baryon asymmetry $B(z=1)\to B(z=\eta^{1/3})\xi\eta^{-1/3}$.
For a given point in parameter space, the mass splitting can thus be adjusted
such that the time when the baryon asymmetry reaches its maximum value coincides
with sphaleron freeze-out. These analytic constraints make it practicable to
identify the maximally possible active-sterile mixings (for normal as
well as inverted hierarchies) that are compared with
presently available experimental and cosmological constraints in \fref{fig:U2maxNHIH}.
Note that when comparing the results presented in 
this article, one should be aware that
\fref{fig:trianglen2ih} and \fref{fig:trianglen2nh}
show the most likely regions of the viable parameter space under the assumption of
the given priors while \fref{fig:U2maxNHIH} shows the total range of viable active-sterile
mixing, what includes also points that have a small posterior likelihood
in the Bayesian analysis. These results are therefore partly complementary
and do not contradict one another.

 \subsection{The next-to-minimal model}
 
 The case with $n_R=3$ has been considered in Refs.~\cite{Akhmedov:1998qx,Drewes:2012ma, Hernandez:2015wna} . It has been shown that successful leptogenesis does not require mass degeneracies in this case, but also the testability of the scenario will be much more challenging due to the enlarged parameter space. 
 
 \subsection{Remark on the reaction rates}
 
 While we have already commented on the agreement in the structure of the networks of fluid
 equations~\eqref{eq:rhonrhonbarav} and the one consisting of Eq.~\eqref{diff:sterile} and
 Eq.~\eqref{evolution:active}, we now also compare the rates that enter the numerical solutions
 based on these two approaches. By definition, the averaged rates $\langle\gamma_N^{(0)}\rangle$ and $\gamma_{\rm av}$
 are identical but slightly different numerical values have been substituted  due to theoretical uncertainties.
 First, $\gamma_{\rm av}$ used in Refs.~\cite{Drewes:2016gmt,Drewes:2016jae} is evaluated with coupling constants at the renormalization scale
 $10^3\,{\rm GeV}$, whereas $\langle\gamma_N^{(0)}\rangle$ from Refs.~\cite{Hernandez:2016kel} takes account of their running at leading logarithmic order.
 Since the running of the ${\rm SU}(2)$ coupling, that dominates these rates, is comparably weak however,
 this does not correspond to the leading discrepancy. Rather, the evaluation of the logarithmically enhanced
 terms due to soft lepton exchange in the $t$-channel for RH neutrino production associated with
 gauge bosons requires a numerical evaluation of
 the scatterings mediated by the one-loop resummed lepton
 propagator in the soft-momentum regime~\cite{Garbrecht:2013urw} or one that is based on analytic approximations~\cite{Anisimov:2010gy,Ghisoiu:2014ena,Garbrecht:2013urw}.
 The value for $\gamma_{\rm av}$ based on Ref.~\cite{Garbrecht:2013urw} therefore turns out to be about
 $30\%$ larger than the value used for $\langle\gamma_N^{(0)}\rangle$ based on Refs.~\cite{Anisimov:2010gy,Ghisoiu:2014ena}.
 
 Further, distinguishing between the averaged rates $\langle\gamma_N^{(0,1,2)}\rangle$ can lead to some quantitative improvement when compared to using a single averaged rate.
However, the discrepancy incurred is within the range
 of the order one uncertainty due to the the missing information
 on the RH neutrino distribution.
 
 Finally, the rates $\alpha^{LC}$ , $\alpha_W^{LC}$ and $\alpha^{LV}$ quoted in \sref{section:LNV}
 include the opening of the phase space due to thermal masses only, as assumed
 in Refs.~\cite{Anisimov:2010gy,Hambye:2016sby,Hambye:2017elz,Ghiglieri:2017gjz}. It has been
 reported that this indeed is a suitable approximation for $\alpha^{LV}$~\cite{Antusch:2017pkq}.
 Phenomenological analyses should therefore include the dominant scattering contributions to $\alpha^{LC}$ and $\alpha_W^{LC}$ as well ({\it cf.} Ref~\cite{Antusch:2017pkq}).

\section{The resonant regime: $L$-violating Higgs-decay leptogenesis}
\label{section:LNV}

As discussed at length above, in the ARS regime the generation of the asymmetry occurs at temperatures much higher than the Majorana masses $M_{Ni}$. Therefore, one can define a \emph{total} lepton number (or fermion number) $L$ as the sum of the SM lepton number and the helicity of the RH neutrinos, in such a way that this is typically treated as approximately conserved. However, as originally argued in~\cite{Hambye:2016sby}, and later confirmed in the Raffelt-Sigl formalism in~\cite{Hambye:2017elz}, leptogenesis can occur also via \emph{L-violating Higgs decays}, i.e.~the decays of the Higgs doublet into a RH neutrino and a SM lepton that do involve a Majorana mass insertion and violate $L$, analogously to leptogenesis from $N$ decays. This LNV effect can become important, or even dominant with respect to the LNC one discussed in detail in the previous sections, in some regions of the parameter space in the resonant regime $|M_{Ni} - M_{Nj}| \lll M_{Ni}$.

\begin{figure}[t]
\centering
\includegraphics[width=15em]{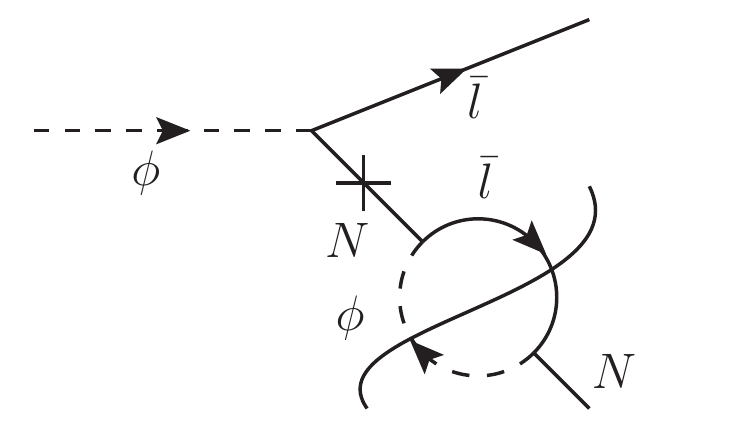}
\caption{Thermal cut in the $\phi \to \bar \ell N$ decay, which gives rise to its purely-thermal $L$-violating CP-violation.\label{fig:decay}}
\end{figure}

In Ref.~\cite{Hambye:2016sby} this phenomenon was studied in the (quantum) Boltzmann equations formalism, with CP-violating rates accounting for the CP violation in the decays $\phi \leftrightarrow \bar \ell N$. This process is depicted in \fref{fig:decay}. Naively, one would not expect any CP violation in this decay: in order to be kinematically accessible, this decay requires $M_N+M_\ell<M_\phi$; in turn, this implies that the propagators in the loop cannot go on shell because of energy conservation, so that no (CP-violating) absorptive part should be present. However, this argument holds true only at zero temperature. Thermal effects yield CP violation in this decay~\cite{Giudice:2003jh,Frossard:2012pc} because the thermal bath can provide the energy required to put on shell the propagators in the loop of Fig.~\ref{fig:decay}. Since also the other Sakharov conditions are fulfilled, as long as the RH neutrino in the final state is out of equilibrium, this process can therefore produce an asymmetry in the early Universe. While this possibility to generate an asymmetry has been known for quite some time~\cite{Giudice:2003jh,Shaposhnikov:2008pf,Drewes:2012ma,Canetti:2012kh,Frossard:2012pc}, this phenomenon was typically considered negligible for the generation of the baryon asymmetry at low scale and therefore neglected, because the corresponding rates have a $M_N^2/T^2$ suppression with respect to the LNC ones discussed in Sec.~\ref{sec:collisionrates}. Instead, as argued recently in~\cite{Hambye:2016sby,Hambye:2017elz}, at the level of the solution to the relevant Boltzmann or density-matrix equations, the parametric dependence of the LNV contribution is different from the LNC one, so that the two phenomena dominate in different regions of the parameter space, as we are going to discuss now.

The LNV and LNC mechanisms can be described in the same framework by means of density-matrix equations similar to the ones discussed in detail in \sref{sec:kinetic}. The overall structure is similar to those, with the main novelty given by the presence of $L$-violating terms, whose rates are suppressed by a factor $M_N^2/T^2$ with respect to the $L$-conserving ones. The detailed form for the $\phi \leftrightarrow \bar \ell N$ processes can be found in~\cite{Hambye:2017elz} (see also~\cite{Canetti:2012kh}). These equations can be solved numerically (in practice if the mass splitting is not too large), or even analytically, in some particular regimes. It is instructive to consider the latter in the weak-washout expansion, limiting to the first non-zero order as in~Eq.~\eqref{eq:as}, where the washout for the LNC ARS mechanism is treated at linear order. 

For the LNC ARS contribution in the linear regime one finds~\cite{Hambye:2017elz,Asaka:2005pn}
\begin{align}
Y^{LNC}_{\Delta B} \ &\simeq  - \, 18.5 \times (\alpha^{LC})^2 \, \alpha^{LC}_W \, \frac{{M_P^*}^{7/3}}{T_c (\Delta M^2_{N 12})^{2/3}} \, (\lambda^\dag \lambda)_{11} (\lambda^\dag \lambda)_{22} \sum_\alpha \delta_\alpha^{LNC} (\lambda \lambda^\dag)_{\alpha \alpha} \;,
\label{LFanalytsol}
\end{align}
where $M_{N1}<M_{N2}$, $T_c \simeq 131.7 \, $GeV is the sphaleron decoupling temperature and the $L$-conserving CP phase is
\begin{equation}
\delta^{LNC}_\alpha \ = \ \frac{\mathrm{Im}\big[\lambda_{\alpha1}^* \lambda_{\alpha2} (\lambda^\dag \lambda)_{21}\big]}{(\lambda^\dag \lambda)_{11} (\lambda^\dag \lambda)_{22}} \;.
\end{equation}
Notice that, as already discussed in Sec.~\ref{sec:cpinvariants}, the LNC contribution in this regime is $\mathcal{O}(\lambda^6)$. The reason is that at $\mathcal{O}(\lambda^4)$ the asymmetry generated vanishes, when summed over SM flavors, since~$\sum_\alpha \delta_\alpha^{LNC}=0$, so that in the LNC case an asymmetric washout of the different flavors is necessary. This yields, in the linear regime, an additional suppression factor $(\lambda \lambda^\dag)_{\alpha \alpha}$ (cf.~also~\eqref{eq:YB}). Equation~\eqref{LFanalytsol} depends on the coefficients 
\begin{equation}
\alpha^{LC} \simeq 3.26 \times 10^{-4} \;, \qquad \alpha_W^{LC} \simeq 1.05 \times 10^{-4}
\end{equation}
that originate from rates of the $L$-conserving production and washout processes $\phi \leftrightarrow \bar \ell N$.

For the LNV contribution one finds~\cite{Hambye:2017elz}, in the same regime,
\begin{align}
Y^{LNV}_{\Delta B} \ &\simeq \ 7.9 \times \alpha^{LC} \, \alpha^{LV} \, \frac{M_P^*}{T_c} \,\frac{M_N^2}{\Delta M^2_{N 12}} \, (\lambda^\dag \lambda)_{11} (\lambda^\dag \lambda)_{22} \, \delta^{LNV} \;,
\label{LVanalytsol}
\end{align}
with coefficient of the $L$-violating production rate
\begin{equation}
\alpha^{LV} \simeq 3.35 \times 10^{-3}
\end{equation}
and $L$-violating CP phase
\begin{equation}
\delta^{LNV} \ = \ \sum_\alpha \frac{\mathrm{Im}\big[\lambda_{\alpha 1}^* \lambda_{\alpha 2} (\lambda^\dag \lambda)_{12}\big]}{(\lambda^\dag \lambda)_{11} (\lambda^\dag \lambda)_{22}}\;.
\end{equation}
Notice that this is nonzero when summed over flavors, so that at leading order in the weak-washout expansion no washout is involved. This implies that the $L$-violating contribution Eq.~\eqref{LVanalytsol} in this regime is $\mathcal{O}(\lambda^4)$, differently from the LNC one. Notice that in the weak-washout regime the two contributions are additive and, since the parametric dependence of Eq.~\eqref{LFanalytsol} and Eq.~\eqref{LVanalytsol} is different, they dominate in different regimes. In particular, for $M \sim 1\,{\rm GeV}$, in the weak-washout regime the $L$-violating contribution can dominate for $|M_{N1}-M_{N2}|/M_{N1} \lsim 10^{-9}$, according to the PMNS and Casas-Ibarra phases.

\begin{figure*}[t]
\includegraphics[width=0.43\textwidth]{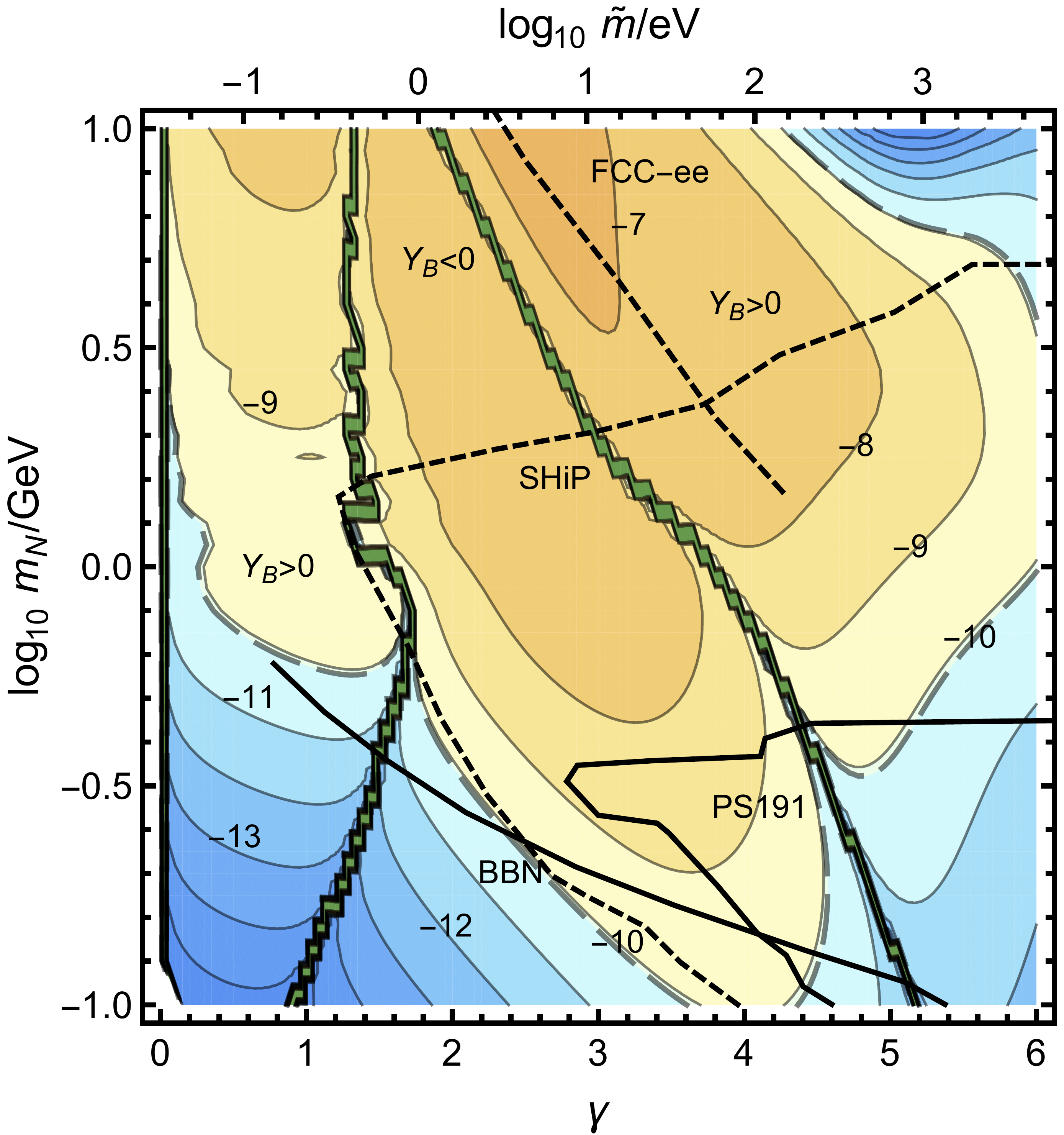}\qquad\qquad
\includegraphics[width=0.43\textwidth]{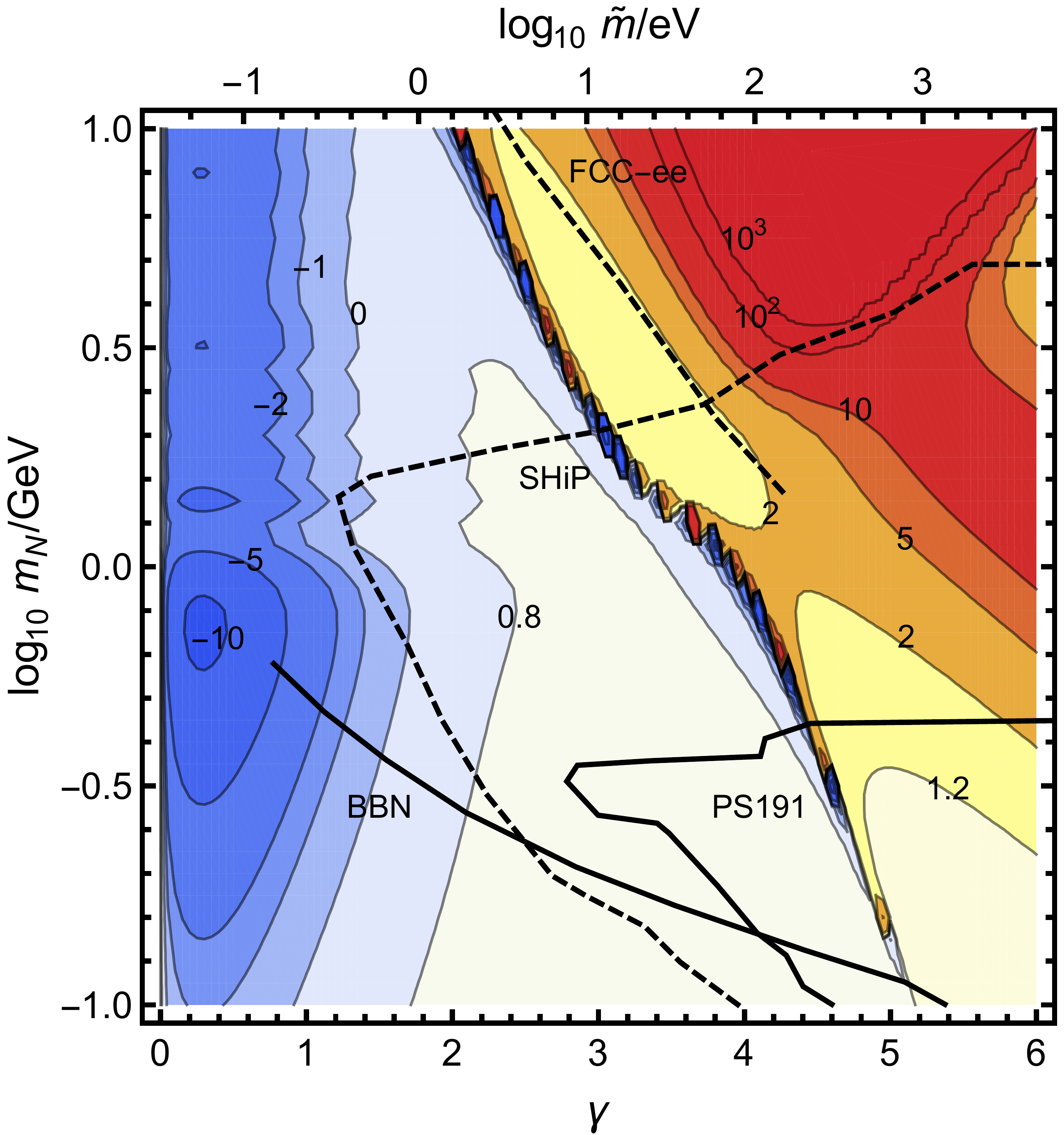}
\caption{Results for 2 RH neutrinos with $(M_{N1}-M_{N2})/M_{N1} = 10^{-10}$ (with $m_N = M_{N1} \simeq M_{N2}$), the PMNS parameters fixed to their best-fit values (with in addition $\delta = - \pi/2$, $\alpha_2=\pi/2$) and the Casas-Ibarra angle fixed to $z = \pi/4 + i \gamma$\label{fig:res_Lviol}. In the left panel, we plot the logarithm base 10 of the $Y_{\Delta B}$ asymmetry obtained. The observed value, $Y_{\Delta B} \simeq 0.86 \times 10^{-10}$ is denoted by the dashed line. In the right panel, we plot the ratio of the full LNC + LNV result to the LNC ARS one only. Exclusion limits and future prospects from various experiments are also shown. For further details, see~\cite{Hambye:2017elz}.}
\end{figure*}

To study the relative importance of the two phenomena beyond the linear weak-washout regime, one can solve numerically the relevant density matrix equations. The numerical results for an illustrative choice of parameters are given in Fig.~\ref{fig:res_Lviol}, with $|M_{N1}-M_{N2}|/M_{N1} = 10^{-10}$. On the left panel we show the region of successful leptogenesis, as given by the full density-matrix equations, which take into account both the LNC and LNV phenomena. On the right panel we show the ratio of the full result with the one obtained artificially switching off the $L$-violating effects. In the light gray region the $L$-violating effects are small, and the ``standard'' LNC ARS phenomenon dominates. Instead, for small values of the imaginary part $\gamma$ of the Casas-Ibarra angle, i.e.~for values of the Yukawa couplings that do not require large cancellations in the seesaw relation, the two phenomena have comparable size (with the LNV one typically dominating for small mass splitting as in Fig.~\ref{fig:res_Lviol}), as already discussed above at the level of the weak-washout analytic solutions, which are valid in this regime. 

In addition, for large values of $\gamma$, the LNV contribution becomes generically dominant, even by many orders of magnitude with respect to the LNC one~\cite{Hambye:2017elz}. The origin of this is clear:  for large $\gamma$ we are in the strong washout regime. In this regime, 
the $L$-conserving washout processes will effectively wash out the LNC part of the asymmetry, whereas the LNV washout part, which is suppressed by an extra $M_N^2/T^2$ factor, will wash out less the asymmetry in the total lepton number $L$ (and only at later times due to this factor), resulting in a dominant $L$-violating part. This dominance occurs generically for Yukawa couplings large enough, even for larger mass splittings. One finds that for $M \sim 1-10$ GeV,  as long as $|M_{N1}-M_{N2}|/M_{N1} \lsim 10^{-5}$, the region in which this occurs corresponds to successful leptogenesis. 

We conclude this section by commenting on the dependence on the initial ``reheating" temperature $T_{in}$, where the RH neutrino population is assumed to vanish. As argued in~\cite{Hambye:2016sby}, the $L$-violating Higgs-decay mechanism is a low-scale one, in the sense that the asymmetry is generated mainly at temperatures close to the sphaleron decoupling one $T_c$. The LNC flavored asymmetries, instead, are mainly produced at $T_{osc} \approx (M_P^* \Delta M_N^2)^{1/3}$ (at least in the weak-washout regime), which easily lies well above $T_c$, in which case the latter is sensitive to physics at scales much higher than the electroweak one.

The discussion in this section takes into account exclusively the baryon asymmetry produced by  $\phi \leftrightarrow \bar \ell N$ processes. As discussed in Sec.~\ref{sec:collisionrates}, a number of other processes are fully relevant, in particular the top-quark and gauge scattering and the IR-enhanced gauge corrections to Higgs decay. These processes will generate a baryon asymmetry in a similar way because they involve as a subprocess the same $\phi \leftrightarrow \bar \ell N$ transition. One could therefore anticipate that, including them, the $L$-violating rates will be enhanced by a factor of few similarly to the $L$-conserving ones. 
A fully quantitative calculation of the baryon asymmetry, with a proper inclusion of all these relevant $L$-conserving and $L$-violating processes,  may therefore modify the numbers given above, but is not expected to change qualitatively the picture.  

\section{Conclusions}

In this Chapter, we have reviewed
the basic mechanism of ARS leptogenesis and beyond that, have put particular emphasis
on the following developments that have taken place during the recent years:
\begin{itemize}
\item
A variety of different parametric regimes and source terms for the ARS scenario that lead to viable leptogenesis
has been found. In addition to the
the original ARS source term,
the weak washout, strong washout and overdamped regimes have been identified and
predictions have been made using numerical as well as analytical methods presented in \sref{section:analytical}.
In categorizing these, weak basis invariants of the neutrino sector have proved
to be valuable, see \sref{sec:cpinvariants}.
\item
There has been an advancement of theoretical computations,
both in refining the operator formalism as well as in formulating and solving
the mechanism in the CTP framework. It has been understood which the relevant
scattering rates for the RH neutrinos are, and state-of-the art computations of these have been
included, see \sref{sec:kinetic}.
\item
Along with the development of these techniques, more accurate and comprehensive
surveys of the parameter space have been performed, both from a Bayesian perspective
as well as by finding the upper limits of the active-sterile mixing imposed by
neutrino phenomenology in conjunction with viable leptogenesis, see \sref{section:numerical}.
\item
Lepton-number violating effects have been included and the regions of parameter space
where these are relevant have been identified, as discussed in \sref{section:LNV}.
\end{itemize}

ARS leptogenesis is a plausible scenario for baryogenesis that will either be further
constrained or even receive
support by  new experimental results in the foreseeable future, see the
accompanying article~\cite{leptogenesis:A05} for more details on phenomenological aspects.
Therefore, it is timely to implement more methodical refinements. Most importantly, uncertainties
due to the momentum-averaging of the RH neutrinos should be removed, LNV effects in the
source and the washout terms should be included for all relevant processes,
%should be combined with the calculations for the standard ARS scenario,
and also the dependence of the rates through the electroweak crossover ought to be accounted for.

Finally, we emphasize that the numerical examples and some of the analytical methods discussed
here are for the case of the SM supplemented with $n_R=2$ RH neutrinos, for definiteness as
well as simplicity. The obvious relaxation $n_R=3$ of this assumption is known to lead
to a largely increased viable parameter space~\cite{Drewes:2012ma,Hernandez:2015wna},
and it will therefore be interesting to investigate ARS leptogenesis in
embeddings of the seesaw mechanism in scenarios beyond the Standard Model.

\section*{Acknowledgments}
This work
has been initiated at the Munich Institute for Astro- and Particle Physics (MIAPP) of the DFG cluster of excellence ``Origin and Structure of the Universe''. The work of DT is supported by a ULB postdoctoral fellowship and the Belgian Federal Science Policy (IAP P7/37).

\bibliography{arsvscp}{}
\bibliographystyle{ws-rv-van}

\end{document}